\documentclass[showpacs,aps,prd,nofootinbib,floatfix,amsmath,amssymb]{revtex4}
\usepackage{graphicx,psfrag}
\usepackage{multirow}
\usepackage{amssymb}
\usepackage[toc,page]{appendix}

\DeclareMathAlphabet{\mathpzc}{OT1}{pzc}{m}{it}
\makeatletter
\newbox\slashbox \setbox\slashbox=\hbox{$/$}
\newbox\Slashbox \setbox\Slashbox=\hbox{\large$/$}
\def\pFMslash#1{\setbox\@tempboxa=\hbox{$#1$}
  \@tempdima=0.5\wd\slashbox \advance\@tempdima 0.5\wd\@tempboxa
  \copy\slashbox \kern-\@tempdima \box\@tempboxa}
\def\pFMSlash#1{\setbox\@tempboxa=\hbox{$#1$}
  \@tempdima=0.5\wd\Slashbox \advance\@tempdima 0.5\wd\@tempboxa
  \copy\Slashbox \kern-\@tempdima \box\@tempboxa}

\def\miss#1{\ifmmode{/\mkern-11mu #1}\else{${/\mkern-11mu #1}$}\fi}
\makeatother
\begin{document}

\title{CP-odd contributions to the  $ZZ^*\gamma$, $ZZ\gamma^*$, and $ZZZ^*$ vertices induced by nondiagonal charged scalar boson couplings}

\author{ A. Moyotl}		
\email[E-mail:]{amoyotl@fcfm.buap.mx}
\author{G. Tavares-Velasco}
\email[E-mail:]{gtv@fcfm.buap.mx}
\author{J. J. Toscano}
\email[E-mail:]{jtoscano@fcfm.buap.mx}
\affiliation{Facultad de
Ciencias F\'\i sico Matem\'aticas, Benem\'erita Universidad
Aut\'onoma de Puebla, Apartado Postal 1152, Puebla, Pue., M\'exico}

\date{\today}

\begin{abstract}
In models with extended scalar sectors with several Higgs multiplets, such as Higgs triplet models, the  $Z$ gauge boson can have nondiagonal couplings to charged Higgs bosons. In a model-independent way, we study the potential contributions arising from such theories to  the CP-violating trilinear neutral gauge boson couplings   $ZZ^*\gamma$, $ZZ\gamma^*$, and $ZZZ^*$, which are parametrized by four form factors: $h^Z_{1,2}$, $f_4^\gamma$, and $f_4^Z$, respectively.  Such  form factors can only be induced if there are at least two nondegenerate charged Higgs scalars and an imaginary phase in the coupling constants. For the masses of the charged scalar bosons we consider values above 300 GeV and   find that the form factors can reach the following orders of magnitude: $|h_1^Z|\sim 10^{-5}-10^{-4}$,  $|h_2^Z|\sim 10^{-7}-10^{-6}$,  $|f_4^\gamma|\sim 10^{-5}-10^{-3}$ and  $|f_4^Z|\sim 10^{-6}-10^{-5}$, though there could be an additional suppression factor arising from the coupling constants. We also find that the form factors  decouple at high energies and are not very sensitive to a change in the masses of the charged scalar bosons. Apart  from a  proportionality factor, our results for the $f_4^Z$ form factor, associated with the $ZZZ^*$ vertex,  is of the same order of magnitude than that induced via nondiagonal neutral scalar boson couplings  in the framework of a two-Higgs doublet model.
\end{abstract}


\date{\today}

\maketitle
\section{Introduction}

Pair production  of electroweak gauge bosons  ($W$, $Z$ and $\gamma$) provides an import test for the gauge sector of the standard model (SM), and opens up the possibility for observing new physics phenomena associated with such particles. In this context, the production of a pair of neutral gauge bosons, $Z\gamma$ or $ZZ$, was studied  at LEP \cite{Achard:2004ds,Abdallah:2007ae}, Tevatron \cite{Abazov:2007wy,Abazov:2007ad,Abazov:2011qp}  and the LHC \cite{Aad:2011xj,Aad:2012awa,Chatrchyan:2012sga,Chatrchyan:2013nda}.  The main production mechanism  proceeds at leading order via the $t$ and $u$ channels, but regardless of the production process, the experimental measurements were found to be consistent with the SM predictions \cite{Achard:2004ds,Abdallah:2007ae,Abazov:2007wy,Abazov:2007ad,Abazov:2011qp,Aad:2011xj,Aad:2012awa,Chatrchyan:2012sga,Chatrchyan:2013nda}. Any deviation in the corresponding cross sections could be a hint of  new physics effects, such as new heavy particles \cite{CortesMaldonado:2011pi}, new interactions \cite{Kober:2007bc,Baur:2000ae}, etc. In particular, $Z\gamma$ and $ZZ$ production could allow us to study the trilinear neutral  gauge boson couplings (TNGBCs) $ZV_iV_j$,  whose framework is best discussed via the effective Lagrangian approach, as in Ref.  \cite{Larios:2000ni}, where  the lowest-dimension effective operators inducing the off-shell TNGBcs were presented. Since the $\gamma\gamma\gamma$ coupling is forbidden by Furry's theorem,  experimentalists have focused their attention on the $Z\gamma\gamma$, $ZZ\gamma$ and $ZZZ$ couplings. On the other hand,  Landau-Yang's theorem forbids any TNGBC with three on-shell gauge bosons, so there are only two vertex functions describing four distinct TNGBCs with one off-shell gauge boson:  $ZV^*\gamma$ and  $ZZV^*$ ($V= \gamma$, $Z$) \cite{Hagiwara:1986vm}. The most general $ZV^*\gamma$ vertex function fulfilling both Lorentz  and electromagnetic gauge invariance is parametrized in terms of four forms factors $h_i^V$ ($i=1,2,3,4$), whereas the $ZZV^*$ vertex function is parametrized by only two forms factors $f_j^V$ ($j=4,5$). While $h_{1,2}^V$ and $f_4^V$ are CP violating,  $h_{3,4}^V$ and $f_5^V$ are CP conserving. Any TNGBC is zero at the tree level in the SM or any  renormalizable extension, only the CP-conserving ones are non-vanishing at the one-loop level in the SM via the fermion triangle \cite{Gounaris:2000tb}, whose contribution is highly suppressed even in the presence of a fourth fermion family  \cite{Hernandez:1999xn}. The respective experimental bounds on TNGBCs are expected to achieve a significant improvement in the LHC  in the forthcoming years. Thus, it is worth studying   any possible contribution to these couplings at  a high energy scale $\Lambda $ via  effective Lagrangian.

Although the CP violation observed in the $K$ meson system can be explained  by the Cabibbo-Kobayashi-Maskawa matrix complex phase, the SM does not predicts enough CP violating effects to explain the current matter-antimatter asymmetry in the universe. Consequently, other sources of CP violation are necessary. Such CP-violating effects may show up via TNGBCs, which are thus worth studying. Along these lines, it was shown that the radiative decay  $Z \to \mu^+\mu^- \gamma$ may be sensitive to both CP-conserving and CP-violating $ZV^*\gamma$ and  $ZZV^*$  TNGBCs. Furthermore, such a process may also be useful to put constraints on the CP-violating forms factors in future linear collider experiments \cite{Perez:2004eb}.  The current experimental bounds on TNGBCs reported by the PDG collaboration \cite{Beringer:1900zz} come from a combination of LEP measurements  \cite{ALEPH:2005aa}, where the experimental results  and the individual analyses are based in reports between 1999 and 2001. However, the L3 collaboration has updated their analyses, resulting in the most restrictive bound on $h_{1,2}^V$ up to date \cite{Achard:2004ds}. On the other hand, a few results have been reported by the CMS \cite{Chatrchyan:2012sga,Chatrchyan:2013nda} and ATLAS collaborations \cite{Aad:2011xj,Aad:2012awa}. The current most stringent  bound on $f_4^V$ was obtained by CMS \cite{Chatrchyan:2012sga}, based on data collected in 2010 and 2011 at $\sqrt{s}=7$ TeV with an integrated luminosity of $5.0 \pm 0.1$ fb${}^{-1}$. A more accurate analysis is expected by both the ATLAS and the CMS collaborations in the forthcoming years. The most stringent bounds on the TNGBCs form factors are shown in Table  \ref{current bound}.

\begin{table}[!htb]
\begin{center}
\begin{tabular}{|c|c|c|}
\hline
Experiment & Limit \\
\hline
\hline
L3  \cite{Achard:2004ds}& $ -0.153<h_1^Z<0.141$ \\
L3  \cite{Achard:2004ds}& $-0{.}087< h_2^Z<0{.}079$\\
CMS \cite{Chatrchyan:2012sga}& $-0{.}011< f_4^Z<0.012$ \\
CMS \cite{Chatrchyan:2012sga}& $-0.013< f_4^\gamma<0.015$\\
\hline
\end{tabular}
\caption{The current most stringent limits  on CP-violating TNGBCs.}
\label{current bound}
\end{center}
\end{table}

Contributions to TNGBCs have been  previously studied in the context of the minimal supersymmetric standard model (MSSM) \cite{Choudhury:2000bw}  and the littlest Higgs model \cite{Dutta:2009nf}, focusing only on   the CP-conserving form factors. Moreover, the CP-violating  $ZZZ^*$ form factors were studied in the framework of two-Higgs doublet model (THDM), where the respective contributions are induced via nondiagonal complex  couplings arising in the neutral scalar sector \cite{Chang:1994cs}. In the framework of several extensions of the SM, flavor change is allowed in the fermion sector via tree level neutral currents. If the respective coupling constants contain an imaginary phase, they can induce  CP violating  TNGBCs at the one-loop level. Another possibility arises when tree level nondiagonal complex couplings $Z \Phi_i^\pm \Phi_j^\mp$ appear  in the scalar sector, with $\Phi^\pm_{i,j}$ charged scalar bosons. In such a case,  non-degenerate charged scalar bosons and a complex mixing matrix are necessary  to induce TNGBCs. In this work we will present an analysis on the CP-violating TNGBCs $ZZ^*\gamma$, $ZZ\gamma^*$ and $ZZZ^*$. In particular we will focus on the one-loop level contributions from nondiagonal complex couplings arising in the charged scalar sector of a SM extension. Our analysis will be rather general as we will use the effective Lagrangian approach. We have organized our presentation as follows. The vertex functions for the TNGBCs and the effective Lagrangian from which they arise are shown in Sec. II. Section III is devoted to the analytical results for the one-loop calculation. Numerical results and discussion are presented in Sec. IV. Finally, the conclusions and outlook are presented in Sec. V.

\section{Trilinear neutral gauge boson couplings}

\begin{figure}[!hbt]
\centering
\includegraphics[scale=0.7]{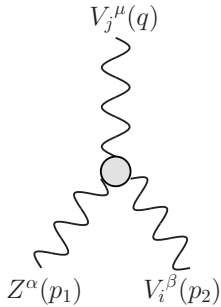}
\caption{Vertex function $ie$ $\Gamma_{ZV_iV_j }^{\alpha\beta\mu}(p_1,p_2,q)$  for the $ZV_iV_j$ ($V_{i,j}=Z$, $\gamma$) couplings. All the four-momenta are outgoing.}
\label{diagramsZVV}
\end{figure}

The most general effective Lagrangian describing TNGBCs $ZV_iV_j$ contains both CP-even and CP-odd terms. Furthermore, if all of the gauge bosons are taken  off-shell, there are both scalar and transverse structures \cite{Gounaris:2000dn}. However, the  scalar terms vanish when on-shell conditions are considered for the $ZV^*\gamma$ and  $ZZV^*$ couplings \cite{Gounaris:2000dn}. Thus, the effective Lagrangians describing such couplings can be written as \cite{Gounaris:1999kf}:

\begin{eqnarray}
{\mathcal L}_{ ZV^*\gamma}&=&  \frac{e}{m_Z^2} \Big{\lbrace} -[h_1^\gamma(\partial^\alpha F_{\alpha\mu})+h_1^Z(\partial^\alpha Z_{\alpha\mu})] Z_\beta F^{\mu\beta} -\frac{1}{m_Z^2} \Big(h_2^\gamma (\partial_\alpha \partial_\beta \partial^\rho F_{\rho\mu}) +h_2^Z [\partial_\alpha \partial_\beta(\partial^2+m_Z^2)Z_\mu] \Big)  Z^\alpha F^{\mu\beta} \label{L-ZVG}
                                        \nonumber\\
{}&{}&-[h_3^\gamma(\partial_\beta F^{\beta\mu})+h_3^Z(\partial_\beta Z^{\beta\mu})] Z^\alpha \tilde{F}_{\mu\alpha} +\frac{1}{2m_Z^2} \Big( h_4^\gamma (\partial^2 \partial^\beta F^{\mu\alpha}) +h_4^Z [(\partial^2+m_Z^2)\partial^\beta Z^{\mu\alpha}] \Big) Z_\beta \tilde{F}_{\mu\alpha}\Big{\rbrace},
                                        \\
{\mathcal L}_{ ZZV^*}&=& \frac{e}{m_Z^2} \Big( -[f_4^\gamma(\partial_\mu F^{\mu\beta})+f_4^Z(\partial_\mu Z^{\mu\beta})] Z_\alpha (\partial^\alpha Z_\beta)+[f_5^\gamma(\partial^\alpha F_{\alpha\mu})+f_5^Z(\partial^\alpha Z_{\alpha\mu})] \tilde{Z}^{\mu\beta} Z_\beta \Big).
\label{L-ZZV}
\end{eqnarray}
Here $\tilde{V}_{\mu\nu}=\epsilon_{\mu\nu\alpha\beta}V^{\alpha\beta}/2$, with $V_{\mu\nu}=\partial_\mu V_\nu-\partial_\nu V_\mu$ standing for the stress tensor of the neutral gauge boson. The operators associated with $h_{2,4}^V$ have dimension eight, but the remaining ones are of dimension six.  While  $h_{1,2}^V$ and $f_4^V$ are CP-odd,  $h_{4,5}^V$ and $f_5^V$ are CP-even. From the effective Lagrangians (\ref{L-ZVG}) and (\ref{L-ZZV}), we obtain the vertex functions $ie\,\Gamma_{ZV_iV_j }^{\alpha\beta\mu}(p_1,p_2,q)$ for the $ZV^*\gamma$ and  $ZZV^*$ couplings respecting Lorentz covariance, $U_{\text{em}}(1)$ gauge invariance, and Bose symmetry, which  are given by \cite{Hagiwara:1986vm,Gounaris:2000tb}:

\begin{eqnarray}
\Gamma_{ZV^*\gamma }^{\alpha\beta\mu}(p_1,p_2,q)&=& i\frac{(p_2^2-m_V^2)}{m_Z^2} \bigg[ h_1^V(q^\beta g^{\alpha\mu}-q^\alpha g^{\beta\mu})+\frac{h_2^V}{m_Z^2} p_2^\alpha\left( (q \cdot p_2)g^{\beta\mu}-q^\beta p_2^\mu\right)
                                        \nonumber\\
{}&{}&-h_3^V \epsilon^{\beta\alpha\mu\rho}p_{2\rho}-\frac{h_4^V}{m_Z^2} p_2^\alpha \epsilon^{\beta\mu\rho\sigma}p_{2\rho}q_\sigma \bigg],\label{ZVgamma}
                                        \\
\Gamma_{ZZV^*}^{\alpha\beta\mu}(p_1,p_2,q)&=&i \frac{(q^2-m_V^2)}{m_Z^2} \left( f_4^V(q^\alpha g^{\mu\beta}+q^\beta g^{\mu\alpha})-f_5^V \epsilon^{\mu\alpha\beta\rho} (p_1-p_2)_\rho\right),\label{ZZV}
\end{eqnarray}
where $m_V$ is the mass of the off-shell $V$ gauge boson, and the overall factor $(p_2^2-m_Z^2)$ in Eq. (\ref{ZVgamma}) is a consequence  of  Bose symmetry. The four-momenta of the gauge bosons are defined in  Figure \ref{diagramsZVV}.

In the next section we will consider a SM extension including complex nondiagonal couplings in the charged scalar sector. We will show that only the CP-odd form factors are induced by this mechanism.

\section{ $ZZ^*\gamma$ and  $ZZV^*$ CP-odd form factors}

We consider a  renormalizable effective theory with several physical charged scalars bosons.\footnote{For instance the low energy effective theory arising from  the 331 model or a Higgs triplet model.}  We will assume that there are trilinear vertices $\Phi_i \Phi_j V$, which can induce TNGBCs at the one-loop level via triangle diagrams.  The Lagrangian describing such trilinear interactions between the neutral gauge bosons and the   charged scalar bosons can be written as ${\mathcal L}={\mathcal L}_{ D}+{\mathcal L}_{ ND}$, where ${\mathcal L}_{ D}$ describes the diagonal couplings and ${\mathcal L}_{ ND}$ the nondiagonal ones. The gauge invariant structure of these Lagrangians is as follows

\begin{eqnarray}
{\mathcal L}_{ D}&=&ie \sum_i Q_\Phi A_\mu \Phi_i^+ \overleftrightarrow{\partial}^{\mu} \Phi_i^-+ig \sum_i g_{ii}^Z Z_\mu \Phi_i^+ \overleftrightarrow{\partial}^{\mu} \Phi_i^-,
\label{L-gammaphiphi}
                                       \\
{\mathcal L}_{ND}&=&i\sum_{i\ne j} g_{ij}^Z Z_\mu \Phi_i^+ \overleftrightarrow{\partial}^{\mu} \Phi_j^-+\text{H.c}.
\label{L-Zphiphi}
\end{eqnarray}
where the coupling constants $g_{ii}^Z$ are real due to the Lagrangian hermicity, whereas $g_{ij}^Z$ could be complex. From here we extract the Feynman rules for the interaction between neutral gauge bosons and the charged scalars bosons, $V_\mu \Phi_i^+(k_1) \Phi_j^-(k_2)$, which has the generic form $ie  g_{ij}^V (k_1-k_2)_\mu$, where all  the four-momenta are outgoing. For $V=\gamma$, there are only diagonal couplings and $g_{ii}^\gamma =Q_\Phi$ is  the charge of the scalar boson in units of $e$.

\begin{figure}[!hbt]
\centering
\includegraphics[width=6.5cm]{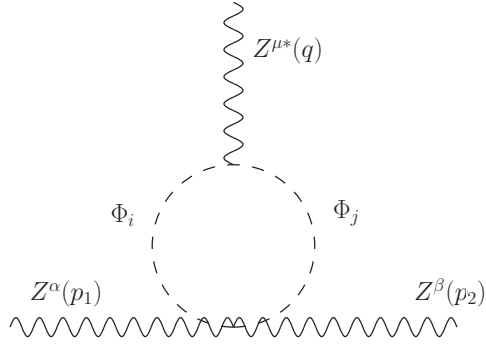}
\caption{Illustrative bubble diagram
 arising from the quartic vertex $ZZ\Phi_i^\pm \Phi_j^\mp$, which can appear in the class of theories we are interested in.  Since this vertex is proportional to the metric tensor $g^{\alpha\beta}$, the amplitude of this diagram is proportional to $q^\mu$ and does not contribute to our CP-violating TNGBC form factors, which neither receive contributions from  all other bubble diagrams obtained by exchanging the $Z$ gauge bosons.}
\label{bubblediagrams}
\end{figure}

It is worth mentioning that in the class of theories we are considering, quartic vertices of the kind $V_i^\alpha V_j^\beta \Phi_i^\pm \Phi_j^\mp$  would also appear.  In principle, this class of vertices can contribute to  TNGBCs at the one-loop level through  bubble diagrams such as the one depicted in Figure \ref{bubblediagrams}. However since the renormalizable quartic  vertex would be proportional to the metric tensor $g^{\alpha\beta}$, the amplitude arising from  this   of diagram can be written as

\begin{equation}
{\cal M}^{\alpha\beta\mu}_{bubble}\sim\int \frac{d^D k}{(2\pi)^D}\frac{(2k+q)^\mu g^{\alpha\beta}}{(k^2-m_i^2)((k+q)^2-m_j^2)}\sim g^{\alpha\beta} q^\mu.
\end{equation}
Therefore this diagram does not contribute to our TNGBC form factors. All other possible diagrams obtained by permuting the $Z$ gauge bosons contribute with terms that can be dropped when the   transversality condition for the on-shell gauge bosons are taken into account and also when considering that the virtual gauge boson is attached to a conserved current.

We now present our results for the one-loop contributions  to  the $ZV^*\gamma$ and $ZZV^*$  couplings, where $V^*$ stands for an off-shell gauge boson. Since only the $Z$ gauge boson  has nondiagonal $Z\Phi_i^\pm\Phi_j^\mp$ couplings, the $Z\gamma^*\gamma$ form factor will not be induced at the one-loop level via this mechanism. Also, no CP-even form factor is induced via scalar couplings since  the Levi-Civitta tensor cannot be generated this way. We will set $Q_\Phi=-1$, but the results will  also be valid for the contribution of a pair of doubly charged scalar bosons, though in this case there will be an additional proportionality factor of $2$ appearing in the  $ZZ^*\gamma$ and $ZZ\gamma^*$ vertices. Once the amplitude for each  Feynman diagram was written down, the Passarino-Veltman method was applied to obtain scalar integrals, which are suitable for numerical evaluation.

\begin{figure}[!hbt]
\centering
\includegraphics[width=10cm]{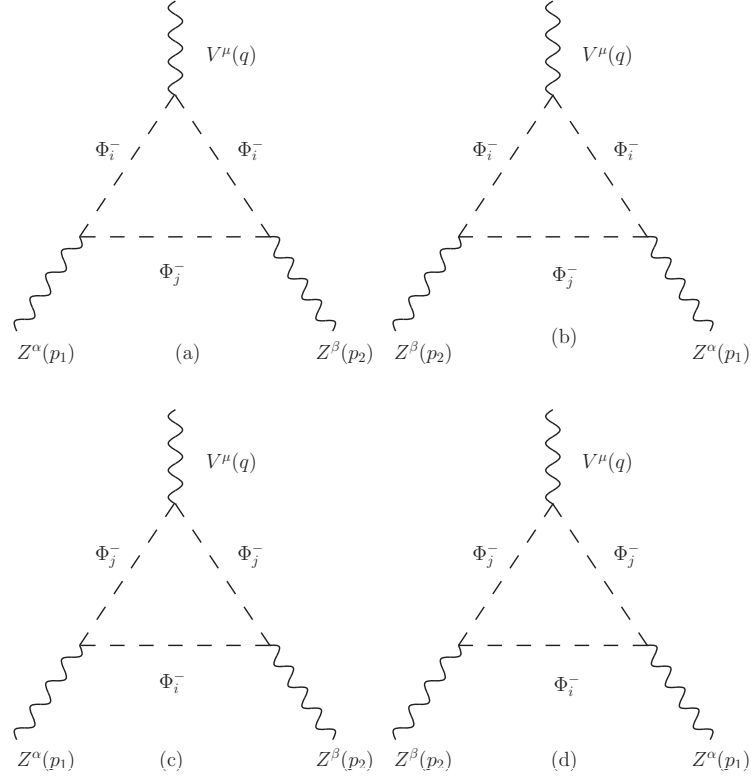}
\caption{One-loop level Feynman diagrams inducing the $ZZ^*\gamma$  ($V=\gamma$ ) coupling.}
\label{ZZVdiagrams}
\end{figure}

\subsection{$Z Z^* \gamma$ coupling}
The Feynman diagrams contributing to the $Z_\alpha(p_1) Z_\beta^* (p_2) \gamma_\mu(q)$ coupling  are shown in Figure \ref{ZZVdiagrams}. It is interesting to note that  diagram (b) is required by  Bose symmetry, whereas  diagrams  (c) and (d), which involve the exchange of the virtual scalars bosons, are necessary to cancel out ultraviolet divergences. After the mass-shell and transversality conditions for the gauge bosons are considered, we obtain the following results

\begin{eqnarray}
\label{h1}
h_1^Z(p_2^2,m_i^2,m_j^2)&=& \frac{m_Z^2 \text{Im}\left(g_{ij}^{Z}{g_{ji}^Z}^*\right)}{6\pi^2(m_Z^2-p_2^2)^3}\Big\{
2 m_i^2
   \left(p_2^2-m_Z^2\right)B_{ii}(0)+3
m_j^2 (m_Z^2-p_2^2)
   \left(m_i^2-m_j^2-m_Z^2\right)   C_{ijj}(p_2^2)\nonumber\\&-&\frac{1}{2}
   (m_i^2-m_j^2)
   \left((m_Z^2 -p_2^2)(B_{ij}(0)-1)+3(m_Z^2 +p_2^2)B_{ij}(p_2^2)-6 m_Z^2
   B_{ij}(m_Z^2)\right)-\left(i\leftrightarrow j\right)
\Big\},\\
h_2^Z(p_2^2,m_i^2,m_j^2)&=& \frac{m_Z^4 \text{Im}\left(g_{ij}^{Z}{g_{ji}^Z}^*\right)}{6\pi^2(m_Z^2-p_2^2)^4}\Big\{
4 m_i^2
 \left(p_2^2-m_Z^2\right)B_{ii}(0)-3 m_j^2
  (m_Z^2-p_2^2)  \left(m_Z^2+p_2^2-2(m_i^2-m_j^2)\right)C_{ijj}(p_2^2) \nonumber\\&-&\frac{1}{2}
  (m_i^2-m_j^2)
 \left(2 (m_Z^2-p_2^2)(B_{ij}(0)-1)+3(m_Z^2+3p_2^2)B_{ij}(p_2^2)-3(3m_Z^2+p_2^2) B_{ij}(m_Z^2)\right)-\left(i\leftrightarrow j\right)
\Big\}\nonumber\\
\label{h2}
\end{eqnarray}
where $p_2$ is the four-momentum of the off-shell $Z$ gauge boson, $m_{i,j}$ stand for the masses of the charged
scalar bosons  and the coefficient $2i\text{Im}(g_{ij}^Z{g_{ji}^Z}^*)=g_{ij}^Z
{g_{ji}^Z}^*-{g_{ij}^Z}^* g_{ji}^Z$ contains the imaginary phase that induces CP violation, which is
necessary to obtain nonzero results and is consistent with the Lorentz structure of the Lagrangian.
We also have introduced the shorthand notation $B_{ab}(c^2)=B_0(c^2,m_a^2,m_b^2)$,
$C_{abc}(p_2^2)=C_0(m_Z^2,0,p_2^2,m_a^2,m_b^2,m_c^2)$ and
$C_{abc}(q^2)=C_0(m_Z^2,m_Z^2,q^2,m_a^2,m_b^2,m_c^2)$, where $B_0$ and $C_0$ stand for
Passarino-Veltman scalar functions. From the above expressions, it is evident that
the form factors vanish when the masses of the charged scalar bosons are degenerate. It is also straightforward to show that  ultraviolet divergences cancel out.

\subsection{$Z Z \gamma^*$ coupling}

The Feynman diagrams for the  $Z_\alpha(p_1) Z_\beta (p_2) \gamma_\mu^*(q)$ couplings are similar to
 those inducing  the $Z Z^* \gamma$ coupling, but in this case the photon is off-shell. The
corresponding form factor is given by:

\begin{eqnarray}
\label{f4gamma}
f_4^\gamma(q^2,m_i^2,m_j^2)&=& \frac{m_Z^2
\text{Im}\left(g_{ij}^{Z}{g_{ji}^Z}^*\right)}{12m_Z^2\pi^2q^2(4m_Z^2-q^2)^2}
\Big\{(m_i^2-m_j^2)
    \left(12 m_Z^2B_{ij}(m_Z^2) +(4m_Z^2-q^2)
(1-B_{ij}(0)) \right)\nonumber\\&+&
\left(q^2 \left(m_i^2+3 m_j^2+7
   m_Z^2-q^2\right)+6
   \left(m_i^2-m_j^2\right)^2-2m_Z^2(8 m_i^2
   -3 m_Z^2)\right)B_{ii}(q^2)\nonumber\\&-&6
   \left(m_i^2-m_j^2-m_Z^2\right)
   \left(m_i^4-2 m_i^2
   \left(m_j^2+m_Z^2\right)+\left(m_j^2-m_Z^2\right)^2+m_j^2 q^2\right)C_{iji}(q^2)
-(i\leftrightarrow j)\Big\},
\end{eqnarray}
where $q$ is  the photon four-momentum. As expected, this form factor is  proportional to
$\text{Im}(g_{ij}^Z{g_{ji}^Z}^*)$ and vanishes when $m_i=m_j$.

\begin{figure}[!hbt]
\centering
\includegraphics[width=10cm]{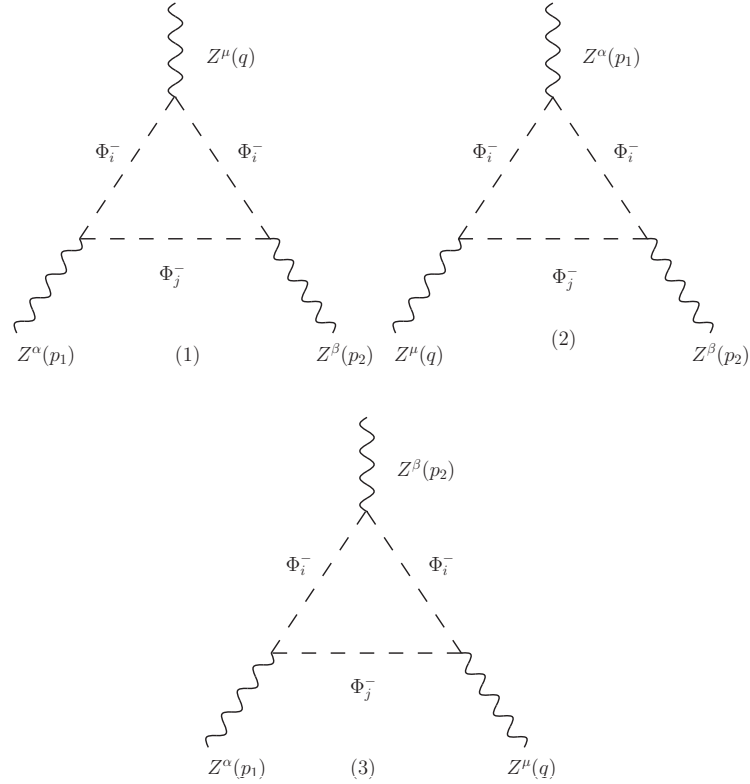}
\caption{Generic one-loop Feynman diagrams for the $ZZZ^*$ coupling. The complete set of diagrams is
obtained by permuting $Z_\alpha(p_1)$ and $Z_\beta (p_2)$ and exchanging the charged scalars bosons.
}
\label{ZZZdiagrams}
\end{figure}

\subsection{$ZZZ^*$ coupling}

Because of Bose symmetry, in the case of the  $Z Z  Z^*$ coupling, there are several more diagrams
than those inducing the $ZZ^*\gamma$ vertex. We  present in Figure \ref{ZZZdiagrams} the  generic
Feynman diagrams from which all diagrams inducing  the $ZZZ^*$ vertex can be generated. Notice that
diagrams (2) and (3)  are obtained from diagram (1) after performing the permutations $Z_\alpha(p_1)
\leftrightarrow Z_\mu (q)$, and $Z_\beta(p_2) \leftrightarrow Z_\mu (q)$, respectively.  Additional
diagrams are obtained from these diagrams following a similar procedure as that described in Fig.
\ref{ZZVdiagrams}: for each one of the Feynman diagrams of Fig. \ref{ZZZdiagrams} there are three more
diagrams that are obtained similarly as  diagrams (b)-(d) of Fig. \ref{ZZVdiagrams}, which  are
obtained from diagram (a)  by permuting the $Z$ gauge bosons and exchanging the charged scalars.
Therefore, there are a total of twelve Feynman diagrams for the $ZZZ^*$ coupling. By using  the
appropriate simplifications, the amplitude of each diagram of  Figure \ref{ZZZdiagrams} reduces to
those of the $Z Z^* \gamma$ and $ZZ \gamma^*$ couplings.  The diagrams (2) and (3) of Fig.
\ref{ZZZdiagrams}  not only are required by Bose symmetry but,  once their amplitudes are added up,
ultraviolet divergences cancel out. After  the Passarino-Veltman method is applied, we obtain the
following result

\begin{eqnarray}
\label{f4Z}
f_4^Z(q^2,m_i^2,m_j^2)&=& \frac{m_Z^2
g_{ii}^Z\text{Im}\left(g_{ij}^{Z}{g_{ji}^Z}^*\right)}{12\pi^2s_Wq^2(q^2-m_Z^2)(q^2-4m_Z^2)^2}\nonumber\\&\times&
\Big\{
q^2 \left(q^2 \left(m_i^2+3 m_j^2+7
   m_Z^2\right)+6
   \left(m_i^2-m_j^2\right)^2-16 m_i^2
   m_Z^2-6 m_Z^4-q^4\right)B_{ii}(q^2)\nonumber\\&+&
   \left(m_Z^2 q^2 \left(10 m_Z^2-13 m_i^2-3
   m_j^2\right)+6 m_Z^2
   (m_i^2-m_j^2)
   \left(m_i^2-m_j^2+2 m_Z^2\right)+q^4
   \left(4
   m_i^2-m_Z^2\right)\right)B_{ii}(m_Z^2)\nonumber\\&-&(m_i^2-m_j^2
   )  \left(2 q^2\left(4 m_Z^2 -q^2\right) (1-B_{ij}(0))-3
   q^4 B_{ij}(q^2) -3m_Z^2 \left(4 m_Z^2-7
q^2\right)B_{ij}(m_Z^2) \right)\nonumber\\&-&6
 \left(m_i^2-m_j^2+2
   m_Z^2\right) \left(m_Z^2 q^2 \left(m_Z^2-3
   m_i^2-m_j^2\right)+m_Z^2
   \left(m_i^2-m_j^2\right)^2+m_i^2
   q^4\right)C_{iij}(q^2)\nonumber\\&-&6 q^2
   \left(m_i^2-m_j^2-m_Z^2\right)
   \left(m_i^4-2 m_i^2
   \left(m_j^2+m_Z^2\right)+\left(m_j^2-m_Z^2\right)^2+m_j^2 q^2\right)C_{iji}(q^2)
-(i\leftrightarrow j)\Big\},
\end{eqnarray}
where $q$ is now the four-momentum of the off-shell $Z$ boson. We note that  all the properties discussed above are also present in this form factor. In the next section we will evaluate the CP-violating TNGBCs for illustrative values of  the charged scalar boson masses and the  four-momentum of the virtual gauge boson.

\section{Numerical results and discussion}
While the diagonal couplings $V\Phi_i^\pm \Phi_i^\mp $ can appear in several extensions of the SM at the tree-level, the
presence of the  nondiagonal couplings  $Z\Phi_i^\pm \Phi_j^\mp $ is less common, but they can be induced
indeed within a more general renormalizable  theory. In order to analyze the CP violating TNGBC form
factors, we will not consider specific values for $g_{ii}^Z$ nor
$\text{Im}(g_{ij}^{Z}{g_{ji}^Z}^*)$. Therefore, the  masses of the charged scalar bosons $m_{i,j}$ and the four-momentum of the virtual gauge boson
will be the only free parameters involved in our analysis.

A particle with the properties of the SM Higgs boson with a mass of 125 GeV has been finally
discovered at the LHC \cite{Aad:2012tfa,Chatrchyan:2012ufa}, and  the search for   singly ($\Phi^+$)
and doubly charged  ($\Phi^{++}$) scalar bosons is still underway by the ATLAS
\cite{Aad:2012rjx,Aad:2013hla,Aad:2012tj,ATLAS:2012hi} and CMS
\cite{Chatrchyan:2012vca,Chatrchyan:2012ya} collaborations. However, no evidence for the existence
of  such scalars bosons has been found up to date.  Based on data collected in 2011, the ATLAS
collaboration  performed  a model independent analysis to search  for a light charged Higgs boson with a mass
in the range 90-160 GeV \cite{Aad:2012rjx}. Independently, the CMS collaboration  reported a
search for the charged Higgs boson of the MSSM with a mass ranging from 80 to 160 GeV
\cite{Chatrchyan:2012vca}. This collaboration   also reported a lower bound on the   doubly charged Higgs boson
mass between 204 and 449 GeV from the processes $pp \to \Phi^{++} \Phi^{--} \to \ell_\alpha^+
\ell_\beta^+ \ell_\gamma^- \ell_\delta^-$, and $pp \to \Phi^{++} \Phi^{-} \to \ell_\alpha^+
\ell_\beta^+ \ell_\gamma^- \nu_\delta$ \cite{Chatrchyan:2012ya}.  Such an analysis was done in the context of  the minimal
type II seesaw model  and the singly and doubly charged scalar bosons were taken to be
mass degenerate.

In the following analysis, we will consider that there is  a  charged scalar
boson with a mass $m_{i}$ above 300 GeV, which is consistent with measurement of the
ATLAS and CMS collaborations. Since the form factors depend on the
splitting between the masses of the charged scalar bosons $\Delta m_{ij}=m_j-m_i$,  we will  use the
parameters $m_i$ and $\Delta m_{ij}$ in our  analysis below, along with the magnitude of the
four-momentum of the virtual gauge boson, which we denote generically as $||p||$. The region of interest corresponds to $\Delta m_{ij}>0$  but for completeness we will also analyze the region $-m_i\le \Delta m_{ij}\le 0$, namely, the scenario with $m_j\le m_i$. Such a  region, which corresponds to a very light charged scalar boson, is not favored by experimental data, but we will consider it in our analysis in order to show the consistency of our results.
 We will thus analyze the behavior of the form factors as functions of  $||p||$ and $\Delta m_{ij}$ for three illustrative  values of  $m_i$. We will show that all the form factors can have both real and imaginary parts. The latter appears when
the magnitude of $||p||$ reaches the value of the sum of the
masses of the charged scalar bosons to which the virtual gauge boson is attached and it is a reflect of the fact that a pair of
real charged scalars could be produced at a collider, rather than two virtual ones,  via an off-shell gauge boson.

\subsection{The  form factor $h_1^Z$}

\begin{figure}[!hbt]
\centering
\includegraphics[width=8cm]{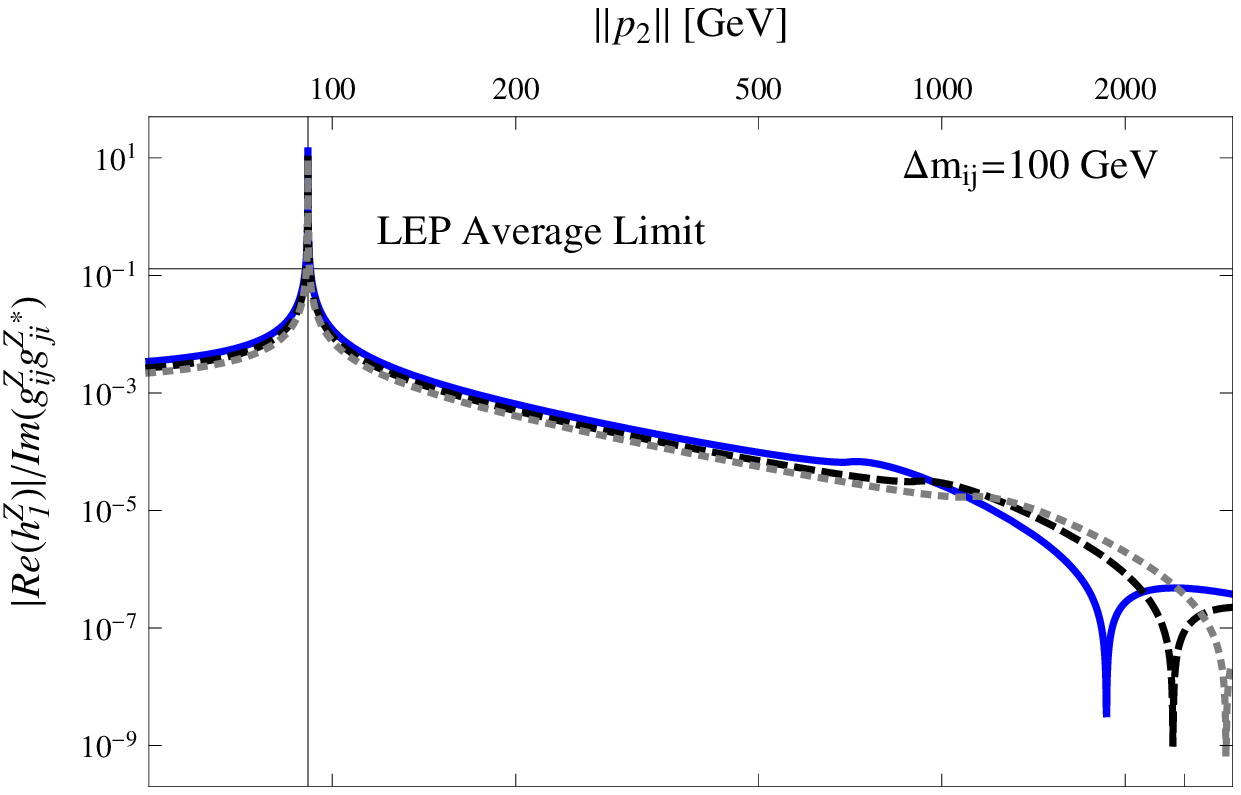}
\includegraphics[width=8cm]{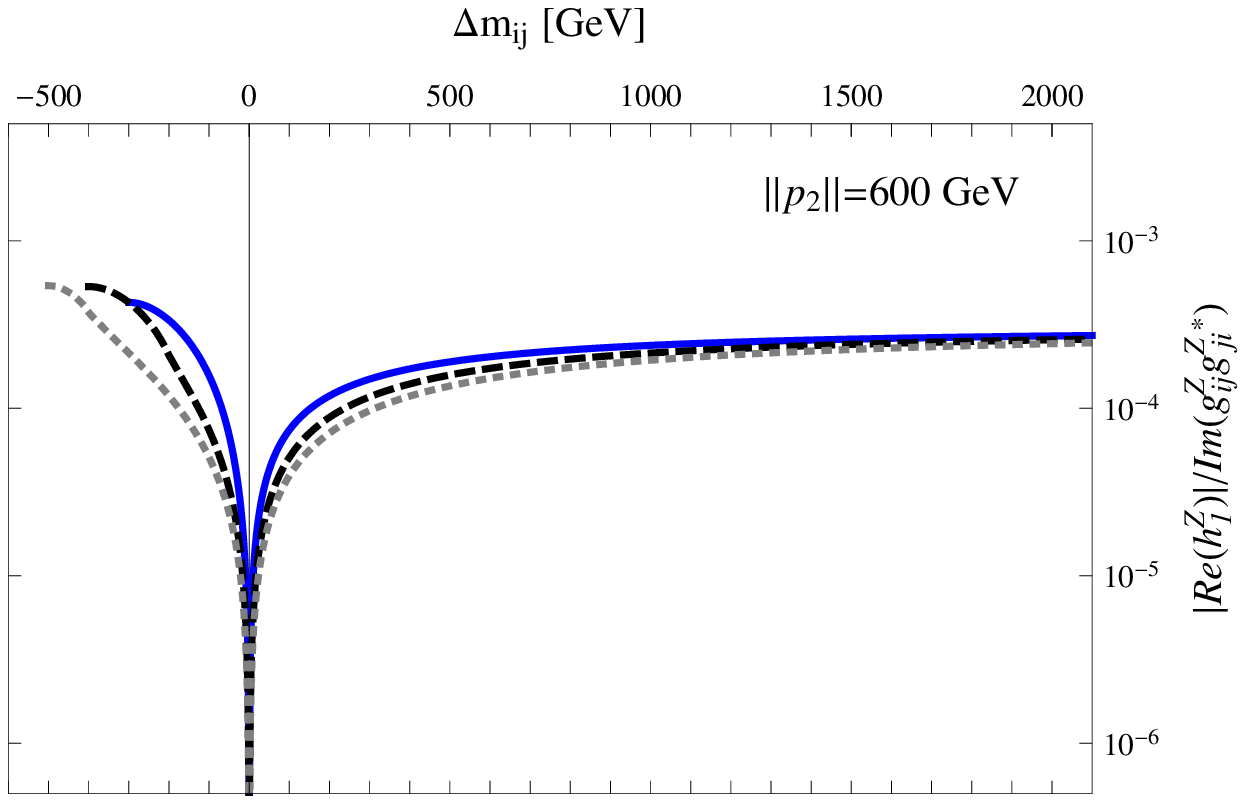}\\
\includegraphics[width=8cm]{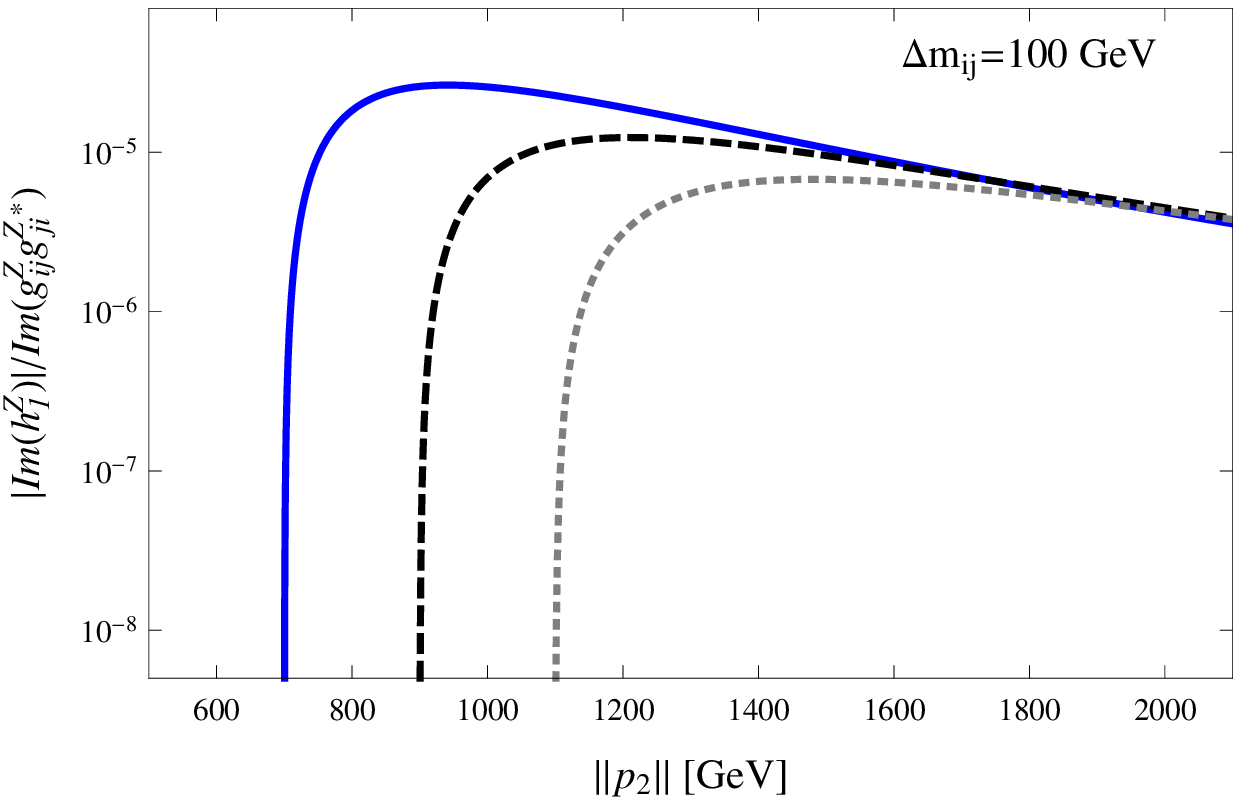}
\includegraphics[width=8cm]{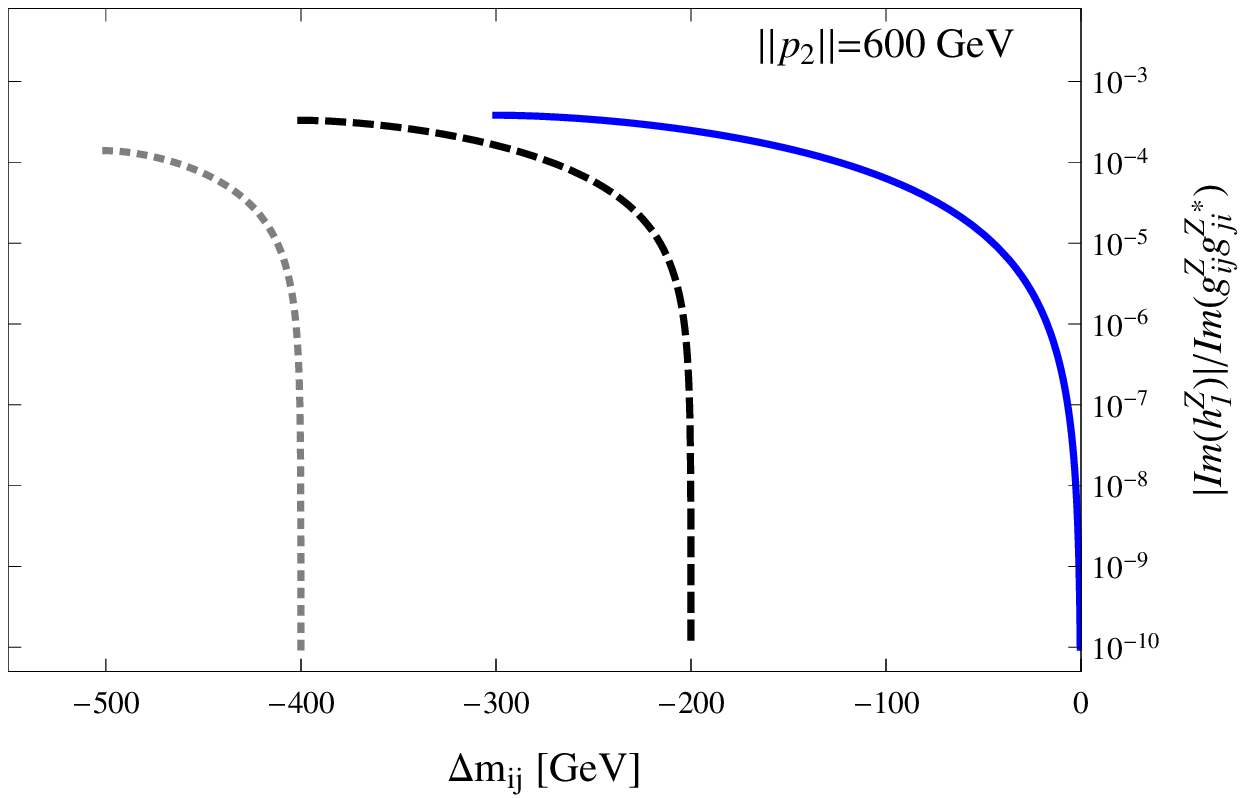}
\caption{Behavior of the real (top plots) and imaginary (bottom plots) parts of the
$h_1^Z$  form factor as functions of the four-momentum magnitude
$||p_2||$   of the virtual $Z$ gauge boson (left plots) and the splitting of the charged
scalar boson masses $\Delta m_{ij}=m_j-m_i$ (right plots). We use the indicated values of $\Delta m_{ij}$ and $||p_2||$ and each curve correspond to a distinct value of the charged scalar mass $m_i$:
$m_i=300$ GeV
(solid line), $400$ GeV (dashed line), and $500$ GeV (dotted line). The horizontal line in the
top-left plot
corresponds to the average value of the LEP lower and upper limits \cite{Achard:2004ds}. We note that  the curves stop because  $m_j=0$ is reached.
}
\label{h1Zplot}
\end{figure}

To begin with, we show in Figure \ref{h1Zplot} the  real (top plots)   and
imaginary (bottom plots) parts of  the $h_1^Z$  form factor as functions of the
four-momentum magnitude $||p_2||$ (left plots) and the splitting of the charged scalar boson masses
$\Delta m_{ij}$ (right plots). For best appreciation of the curves
we show the absolute value of the real and imaginary  parts of the form factor.  We use three distinct values of $m_i$: $300$, $400$ and $500$
GeV.

We
will first analyze the behavior of the real part  of the $h_1^Z$ form factor (top plots). It can be
observed in Eq. (\ref{h1}) that $h_{1}^Z$ includes the term
$m_Z^2-p_2^2$ in the denominator and thus it becomes undefined when $p_2^2\to m_Z^2$, which
explains the sharp peak at $||p_2||=m_Z$ in the top-left plot, where we include a vertical line for illustrative purposes.
This effect is not in conflict with Landau-Yang's theorem since the full vertex function
(\ref{ZVgamma}) does vanish when all the $Z$ gauge bosons are taken on-shell.
In the $||p_2||>1600$ GeV region of the top-left plot a dip appears in each curve, though in the plot it is only visible in the  $m_i=300$ GeV curve. This is
due to a flip of sign of $h_1^Z$:  in the $||p_2||<m_Z$  region the  $h_1^Z$ form
factor is positive, whereas in the $||p_2||>m_Z$ region it is negative and changes sign again in the dip located at $||p_2||>1600$.  It is also
interesting to note that the largest
values of the real part of $h_1^Z$ are reached  around $||p_2||\sim m_Z$.  For instance, in units of
$\text{Im}(g_{ij}^{Z}{g_{ji}^Z}^*)$, we have $|h_1^Z|\sim 10^{-3} $ for $||p_2||\simeq 100$ GeV and
$|h_1^Z|\sim 10^{-5}$   for $||p_2||=1000$ GeV. This is to be compared
with  the average value obtained from the LEP lower and upper bounds on the $h_1^Z$ form factor
\cite{Achard:2004ds} (horizontal line of the top-left plot).
 As far as the behavior of
$h_1^Z$ as a function of $\Delta m_{ij}$ is concerned, while the region
with large  $\Delta m_{ij}$ corresponds to a  heavy $m_j$,  the $-m_i\le \Delta
m_{ij} \le 0$ region corresponds to the scenario with a very light charged scalar boson with a
mass lying in the interval $0\le m_j\le m_i$. Although this scenario seems to be ruled out by
experimental data, we include it in our analysis for completeness. We note that in these and subsequent plots the stop of the curves is due to the reach of $m_j=0$. As expected, since degenerate
charged scalar bosons
do not give rise to CP-violating form factors, the form factors vanish when $\Delta m_{ij}=0$, this is why in the top-right plot we observe a sharp dip at $\Delta m_{ij}=0$,
which is due to the vanishing of the form factor.   We
also note that the real part of the $h_1^Z$ form factor is not sensitive to a change in the
splitting $\Delta m_{ij}$, so the different curves are almost indistinguishable.

As far as the imaginary part of $h_1^Z$ is concerned, which we show in the bottom plots of Fig.
\ref{h1Zplot}, since the photon must couple to the same charged scalar boson circulating into the loop, the $Z$ gauge boson must necessarily couple to distinct charged scalar
bosons. Thus  the
imaginary part of $h_1^Z$ can only appear when $||p_2||\ge m_i+m_j$, which is evident in the
curves of the bottom-left plot. It means that a  higher energy would be required to measure such
imaginary part unless there was a relatively light charged scalar, which is a scenario ruled out by
experimental data. On the other hand, when the value of $||p_2||$ is fixed  the
imaginary part of $h_1^Z$ can only appear in the small region $0\le m_j\le ||p_2||-m_i$ or $-m_i\le \Delta
m_{ij}\le
||p_2||-2m_i$, as shown in the bottom-right plot. This interval becomes narrower for increasing $m_i$: for instance when $||p_2||=600$ GeV,   the imaginary part of $h_1^Z$ is nonvanishing in the interval $0\le m_j\le 300$ GeV for $m_i=300$ GeV,  $0\le m_j\le 200$ GeV for $m_i=400$ GeV and $0\le m_j\le 100$ GeV for $m_i=500$ GeV. Again, we only include these results in
our analysis for completeness.
In general terms, we observe that the imaginary part of the  $h_1^Z$ form factor can have a
size  of
similar order of magnitude than its real part, in the same interval of the region of parameters where the former is nonvanishing. However the
maximal size of the imaginary part is reached for a very light charged scalar, whereas the
maximal size of the real part can be reached around the $Z$ resonance, where there is no imaginary
part.

\subsection{The  form factor $h_2^Z$}
\begin{figure}[!hbt]
\centering
\includegraphics[width=8cm]{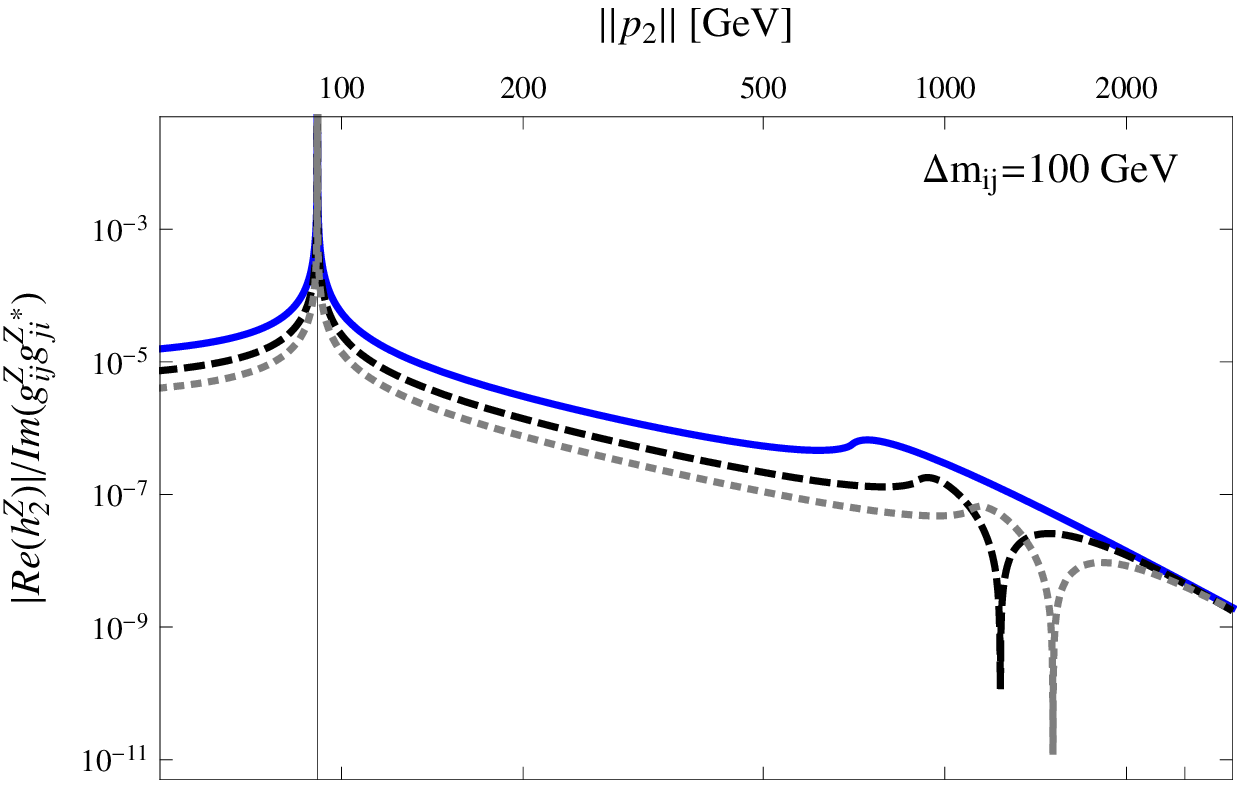}
\includegraphics[width=8cm]{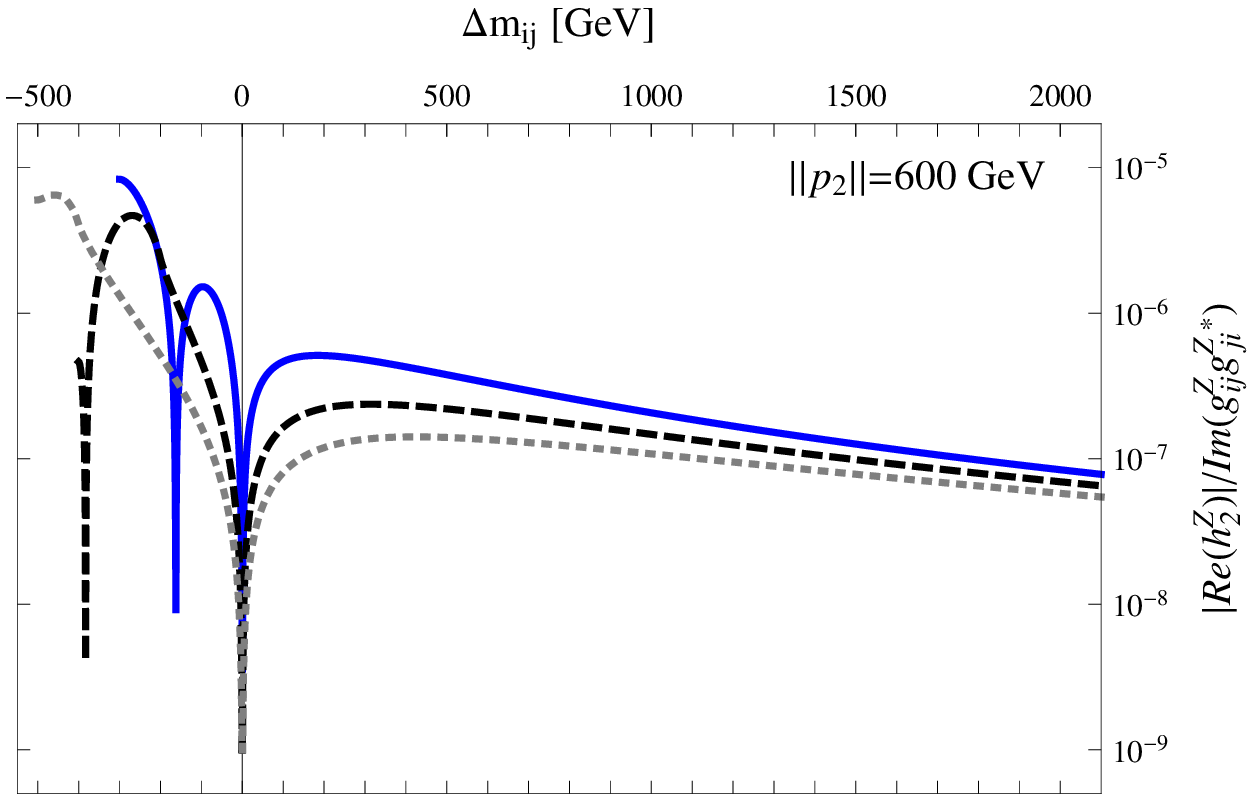}\\
\vspace{-0.cm}
\includegraphics[width=8cm]{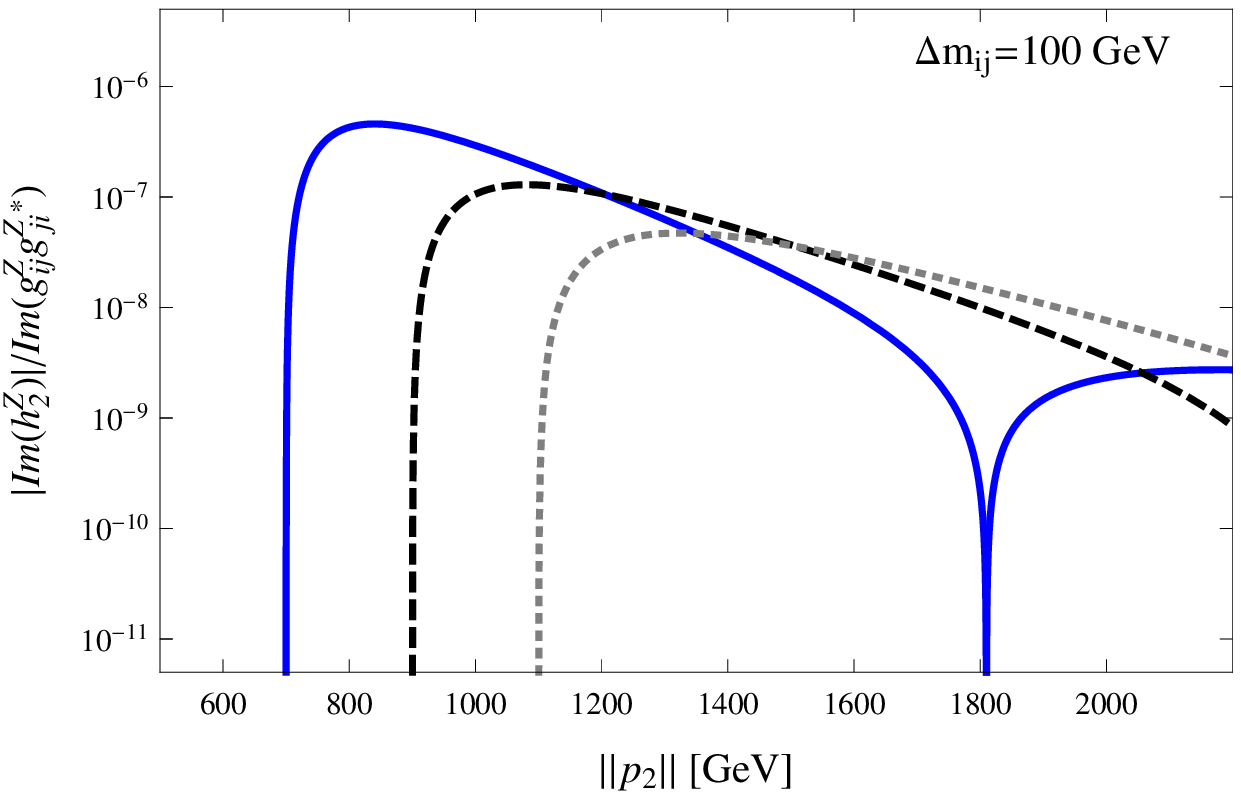}
\includegraphics[width=8cm]{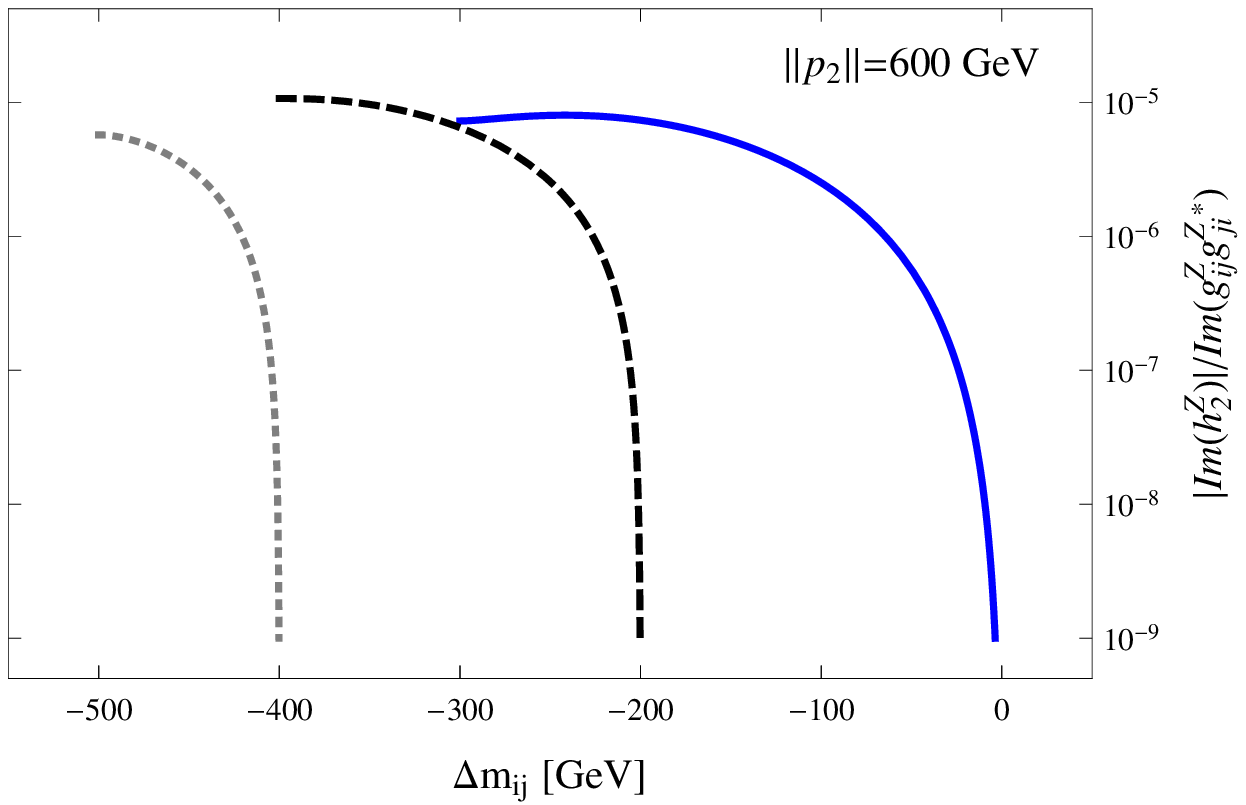}
\caption{
The same as in Fig. \ref{h1Zplot}, but for the $h_2^Z$ form factor.}
\label{h2Zplot}
\end{figure}

We now  present in Figure \ref{h2Zplot} the corresponding plots for the $h_2^Z$ form factor, namely, we
show the  behavior of the real (top plots)  and imaginary
(bottom plots) parts  of the form factor $h_2^Z$ as functions
of $||p_2||$ (left plots) and
$\Delta m_{ij}$ (right plots).  We use the same set of parameters as in Fig. \ref{h1Zplot}.
We note that in general the $h_2^Z$ form factor shows a similar behavior to that of the $h_1^Z$ form
factor, though there are some slight differences. We first analyze the top plots of Fig.
\ref{h2Zplot}, which show the real part of $h_2^Z$.
We observe that apart from the sharp peak at $||p_2||=m_Z$ in the top-left plot,  there are also  dips at high energy, such as occurs in the respective
$h_1^Z$ curves. Such dips, which are a result of the flip of sign of the form factor,  are now shifted
to the left and they appear at a higher $||p_2||$ for a smaller $m_i$, which is opposite to the behavior of $h_1^Z$. Thus the dip for $m_i=300$
GeV curve is not shown in the plot. In the $||p_2||<m_Z$ region
$h_2^Z$ is negative, whereas in the  $||p_2||>m_Z$ region it is positive, contrary to what happens
with $h_1^Z$. After the dip at $||p_2||\ge 1000$ GeV, $h_2^Z$ becomes negative again.
We also note that  the magnitude of $h_1^Z$ is greater than that of $h_2^Z$, although for very high
values
of $||p_2||$ or very heavy $m_{i}$, the both $h_1^Z$ and $h_2^Z$ are considerably suppressed.
As expected, the largest values of the real part of $h_2^Z$ are reached around $||p_2||\sim m_Z$.
For instance, in units of
$\text{Im}(g_{ij}^{Z}{g_{ji}^Z}^*)$, we have $|h_2^Z|\sim 10^{-6} $ for $||p_2||=100$ GeV and
$|h_2^Z|\sim 10^{-7}$  for $||p_2||=1000$ GeV.  It can also be observed
that, $h_2^Z$ is more sensitive than $h_1^Z$ to a change in the value of $m_i$. Regarding the
top-right plot, again we observe that the $h_2^Z$ form factor vanishes when $m_i=m_j$, which shows
the consistency of our result. The curves of the
top-right plot show a dip in the $-m_i\le \Delta m_{ij}\le 0$ region , which are due to a change of sign and do not appear in the case of the $h_1^Z$
form factor. Although $h_2^Z$ can reach its largest values in this region,   as explained above, it
corresponds to the case of a very light charged scalar.

As in the case of $h_1^Z$ (both $h_1^2$ and $h_2^Z$ form factors arise from the same vertex
function), the imaginary part of $h_2^Z$ would be nonvanishing when $||p_2||\ge m_i+m_j$, which
again  is evident in the bottom-left plot of
Figure \ref{h2Zplot}.  Therefore, to observe such an imaginary part a higher
energy   than that required to
observe the respective real part would be required. Furthermore, although the sizes of the imaginary part and the real
part are similar,  the real part  can reach larger
values in the interval where the imaginary part  vanishes. Thus, in general the
imaginary part  is smaller than the largest possible values of the
real part, which is reached around the $Z$ resonance. As far as the behavior of the imaginary part
of $h_2^Z$ as a function of $\Delta m_{ij}$ is concerned,  as expected it is nonvanishing in the  interval $-m_i\le \Delta
m_{ij}\le
||p_2||-2m_i$, where   $m_j$ is  very light. In this region the imaginary part of $h_2^Z$   can reach values slightly larger than the real part.

\subsection{The form factor $f_4^\gamma$}

\begin{figure}[!hbt]
\centering
\includegraphics[width=8cm]{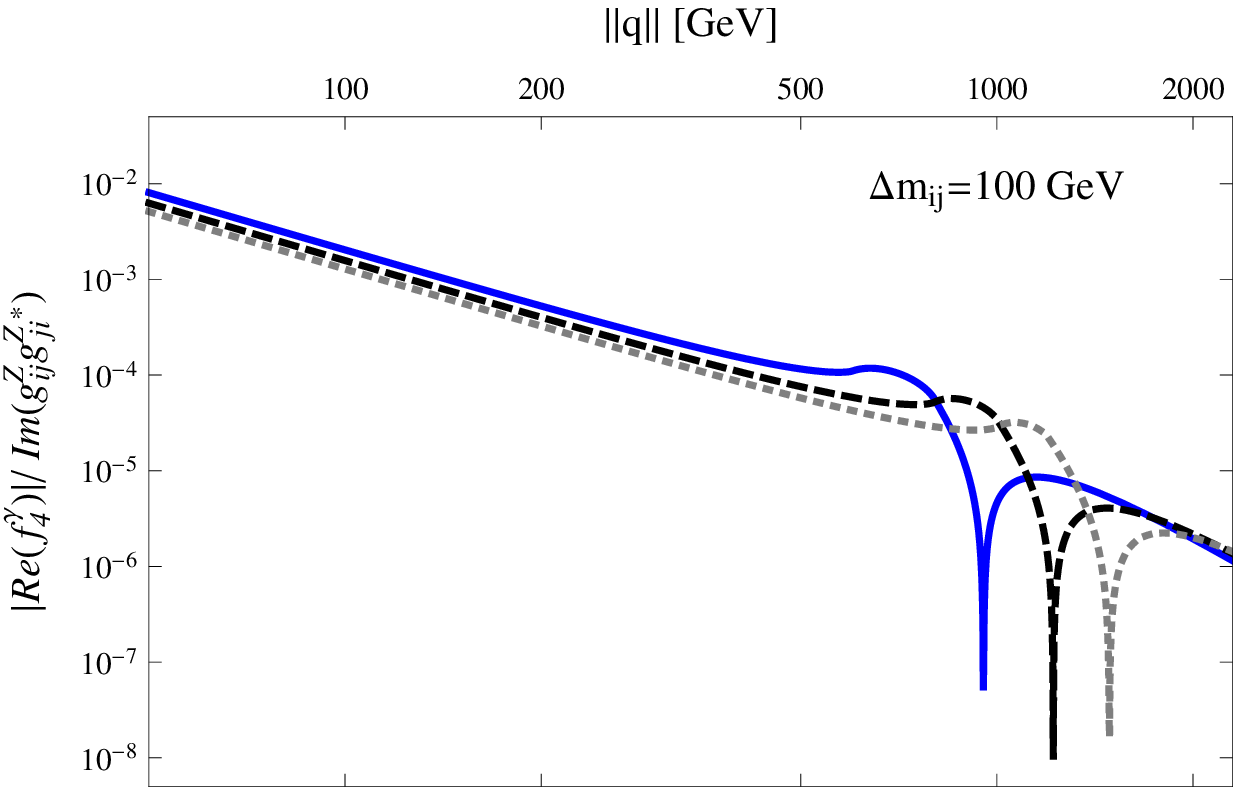}
\includegraphics[width=8cm]{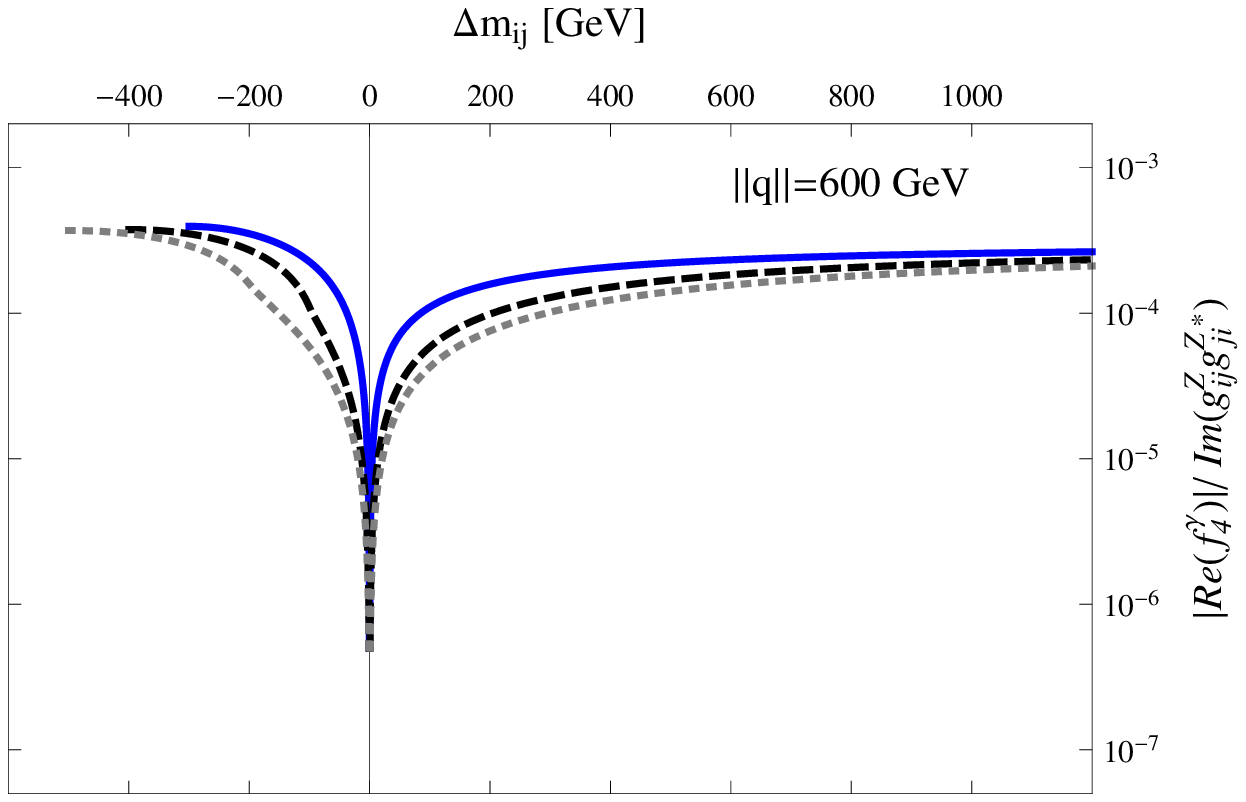}\\
\vspace{-0.cm}
\includegraphics[width=8cm]{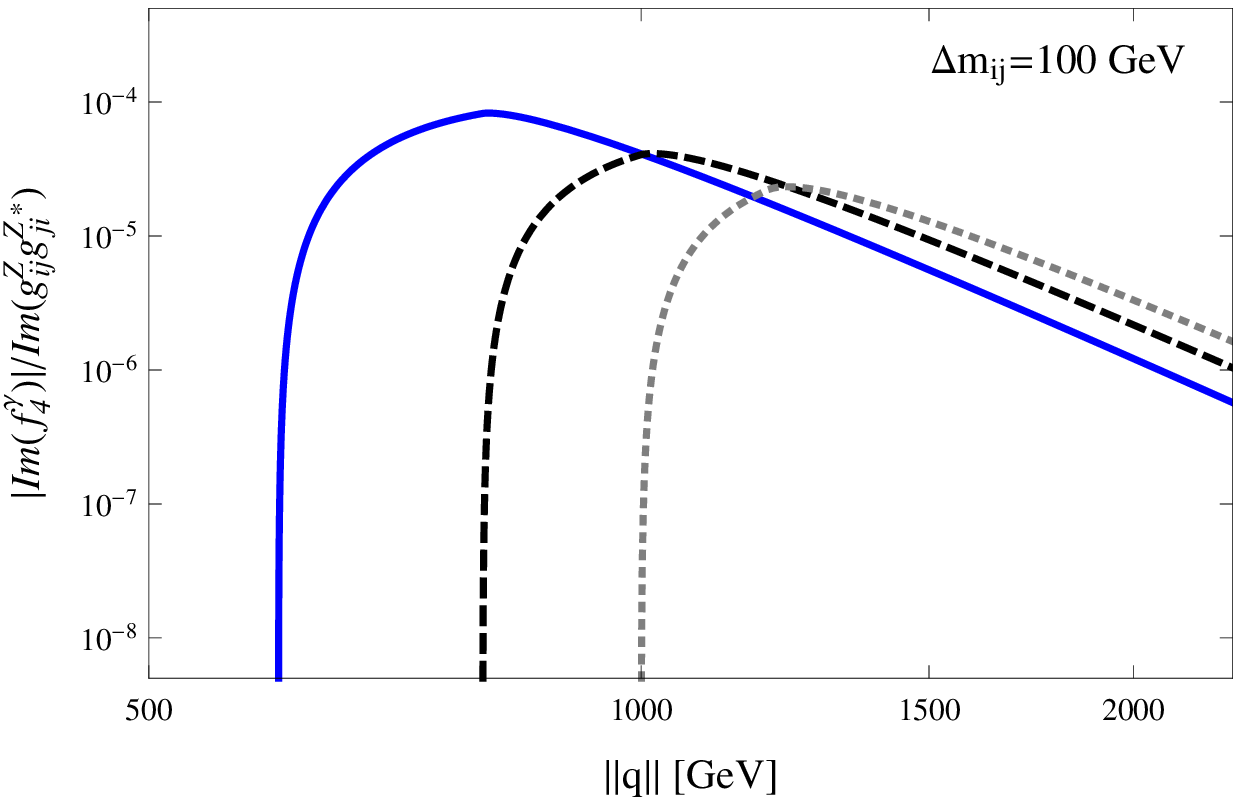}
\includegraphics[width=8cm]{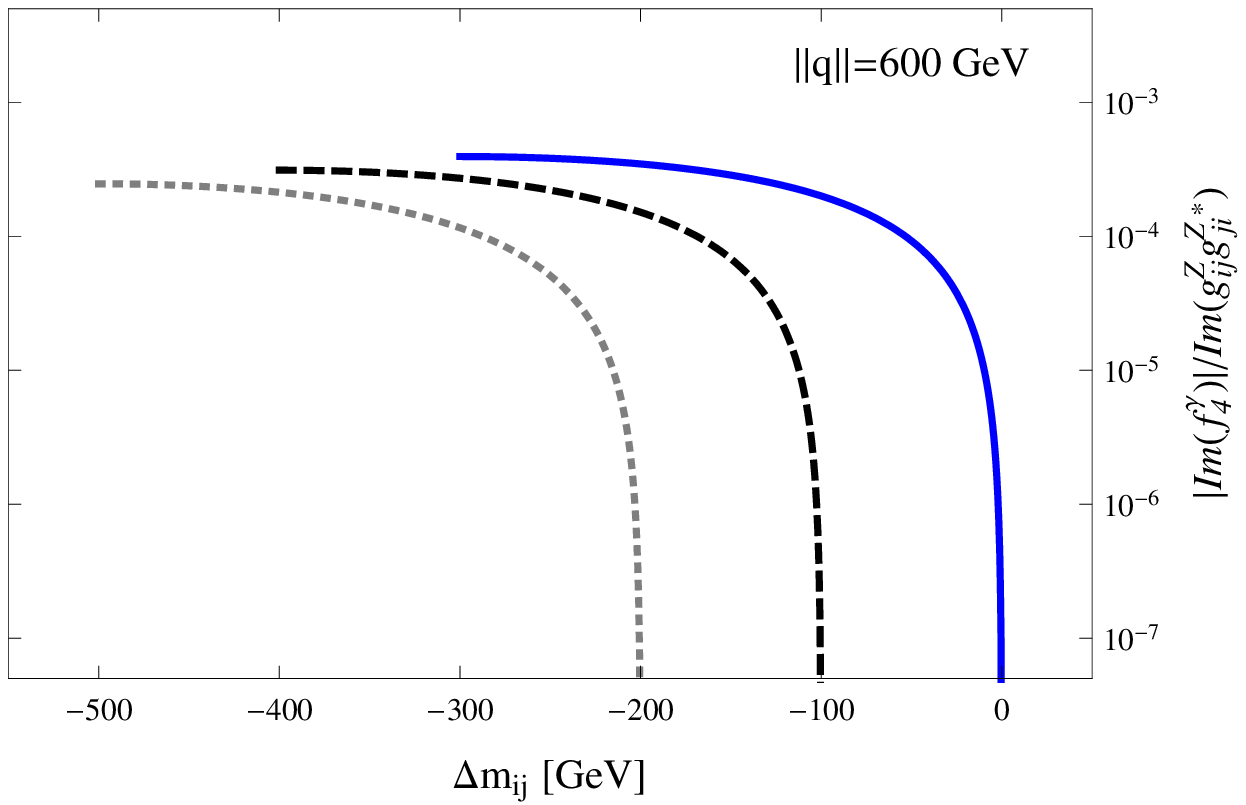}
\caption{Behavior of the real (top plots) and imaginary (bottom plots) parts of the $f_4^\gamma$  form factor as a function of the four-momentum magnitude $||q||$of the virtual photon  (left plots) and  the
splitting of the charged scalar boson masses $\Delta m_{ij}$ (right plots).
 We use the indicated values for $\Delta m_{ij}$ and $||q||$ and each curve correspond to a distinct value of the charged scalar mass $m_i$:
$m_i=300$ GeV
(solid line), $400$ GeV (dashed line), and $500$ GeV (dotted line). As explained in the text, the imaginary part of $f_4^\gamma$  arises only when $||q||\ge 2m_i$ in the bottom-left plot and $0\le m_j\le \frac{||q||}{2}$ or $-m_i \le \Delta m_{ij}\le
\frac{||q||}{2}-m_i$ in the bottom-right plot.}
\label{f4g}
\end{figure}

We now turn to analyze the $f_4^\gamma$ form factor. We note that it becomes
undefined when $||q||\to 0$ as can be inferred from Eq. (\ref{f4gamma}).
Nevertheless, Landau-Yang's theorem is obeyed as can be deduced after a closer inspection of Eq.
(\ref{ZZV}).   In the top plots of  Figure \ref{f4g} we
show the real part of  $f_4^\gamma$ as a function of the four-momentum magnitude $||q||$  (left
plot) and the mass splitting $\Delta m_{ij}$ (right plot). On the
other hand, similar plots are shown in the bottom part of the panel that illustrate the behavior of the imaginary part of $f_4^\gamma$. In this analysis we will use the same set of $m_i$  values  used in Fig.
\ref{h1Zplot}.

As far as the real part of the $f_4^\gamma$ form factor is concerned, we note that the dip
appearing in each curve of the top-left plot corresponds to a flip of sign of $f_4^\gamma$, which
turns from negative to positive at $||q||\simeq$ 950, 1100, and 1500  GeV for $m_i=$300, 400, and
500 GeV, respectively. We also observe that this form factor decouples at high energy,  where it has a negligible magnitude, but in
the interval between
$100$ GeV and 900 GeV  it can reach values as higher as $10^{-5}-10^{-3}$, in units of $\text{Im}(g_{ij}^{Z}{g_{ji}^Z}^*)$. It is also interesting to note
that the real part of $f_4^\gamma$  is not very sensitive to a variation of the charged scalar mass
$m_i$, which contrasts with the behavior of the imaginary part.  On the other hand, in the top-right plot of Figure \ref{f4g}, we  show the behavior of
$f_4^\gamma$ as a function of $\Delta m_{ij}$ for
$||q||=600$ GeV and the  same three values of $m_i$ used in previous analyses. We observe
that the magnitude of $f_4^\gamma$   does not increase significantly as   $\Delta m_{ij}$ increases.
As expected, this form factor vanishes when $\Delta m_{ij}=0$. We note that the maximal values of
the real part of $f_4^\gamma$ are of the order of $10^{-3}-10^{-2}$ times the factor $\text{Im}(g_{ij}^{Z}{g_{ji}^Z}^*)$.

In the bottom plots of Figure \ref{f4g} we show the behavior of the imaginary part of $f_4^\gamma$.
Since the virtual photon must necessarily couple to the same
charged scalar boson, the imaginary part of this form factor would  arise
when $||q||\ge\min \left(2m_i,2m_j\right)$, which is evident in the bottom-right plot, where the imaginary part is nonzero for $||q||\ge 2m_i$.
For fixed $m_i$, the interval for nonvanishing imaginary part can be written as $-m_i \le \Delta
m_{ij}\le \frac{||q||}{2}-m_i$ or $0\le m_j\le \frac{||q||}{2}$. This is the reason why
 the curves in the  bottom-right plot  are nonvanishing only in the small region where $-300$ GeV $\le \Delta m_{ij}\le 0$ GeV for $m_i=300$ GeV,
 $-400$ GeV $\le \Delta m_{ij}\le -100$ GeV for $m_i=400$ GeV and $-500$ GeV $\le \Delta m_{ij}\le -200$ GeV for $m_i=500$ GeV.
Again we note that the imaginary part of $f_4^\gamma$  has a similar order of magnitude than its
real part, namely $10^{-5}$ times $\text{Im}(g_{ij}^{Z}{g_{ji}^Z}^*)$. However, as in the previous form factors, the real part can reach larger values in the region where the imaginary part vanishes than in the region where both of them are nonzero.

\subsection{The form factor $f_4^Z$}

\begin{figure}[!hbt]
\centering
\includegraphics[width=8cm]{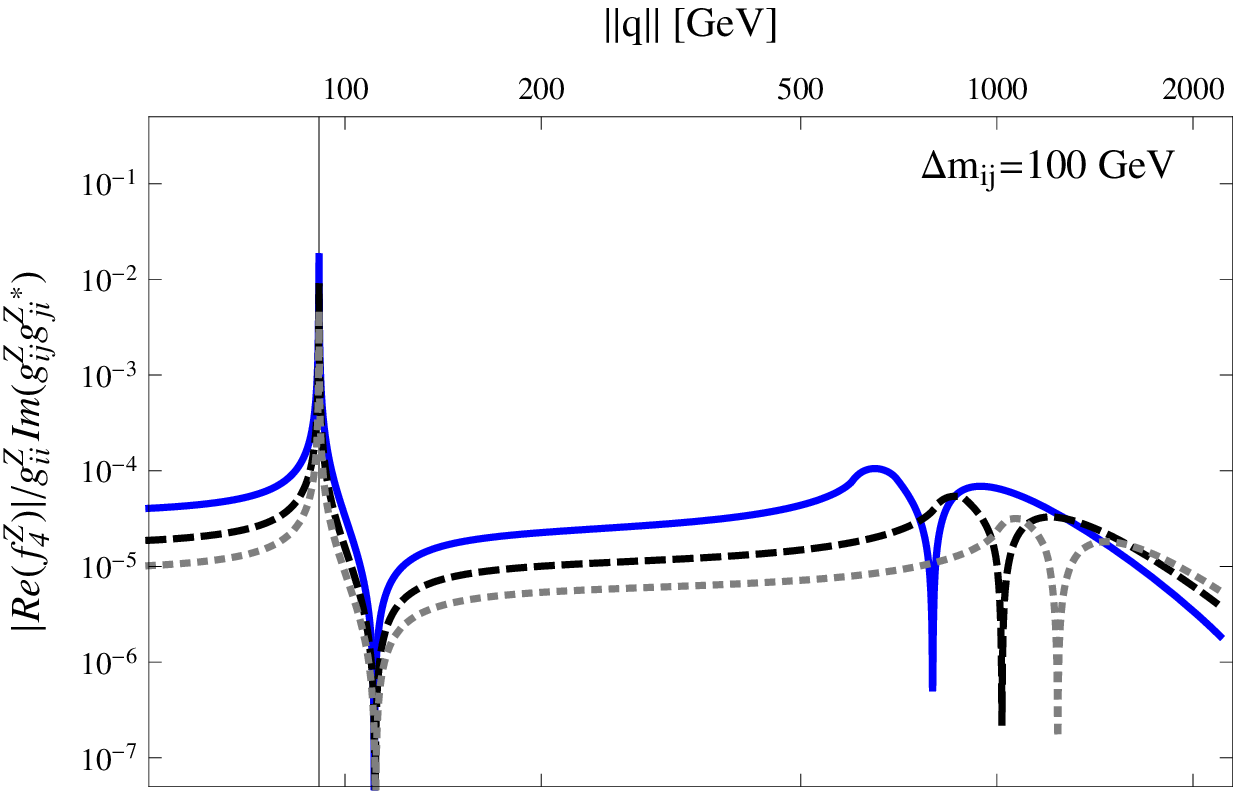}
\includegraphics[width=8cm]{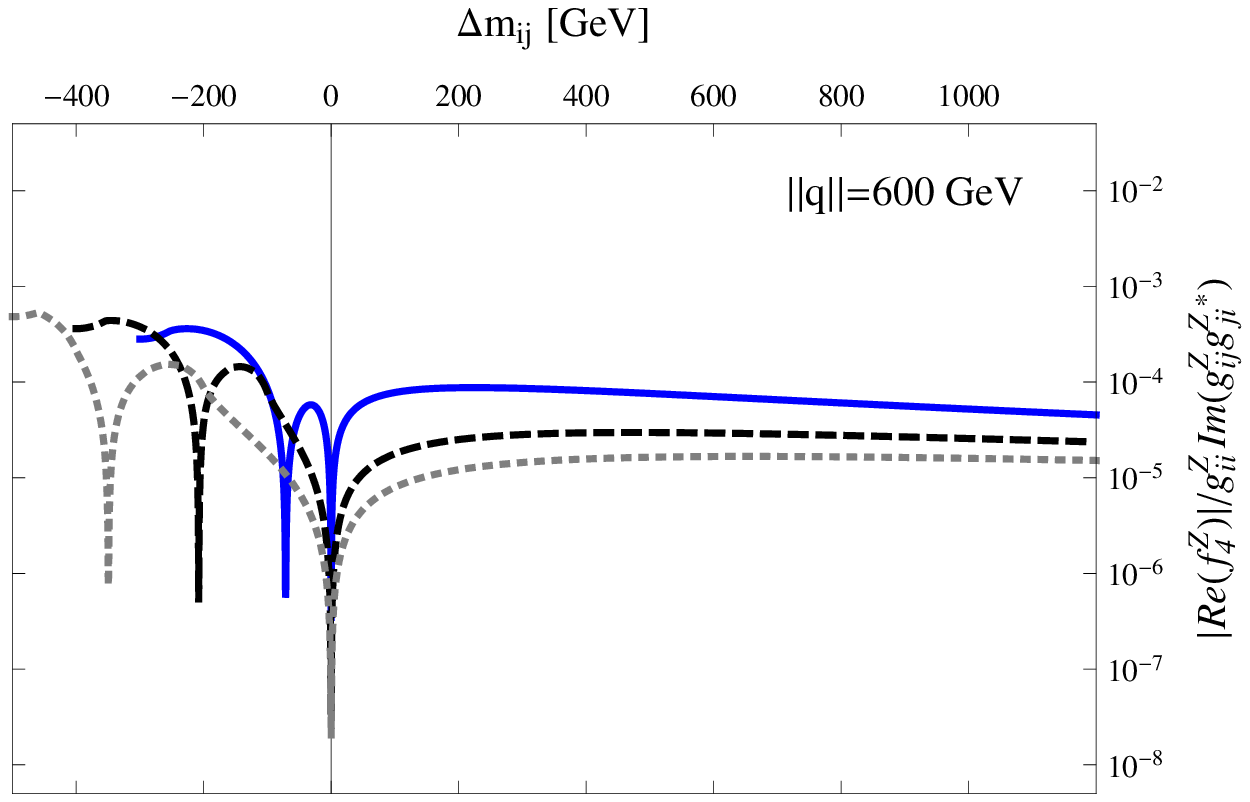}\\
\vspace{-0.cm}
\includegraphics[width=8cm]{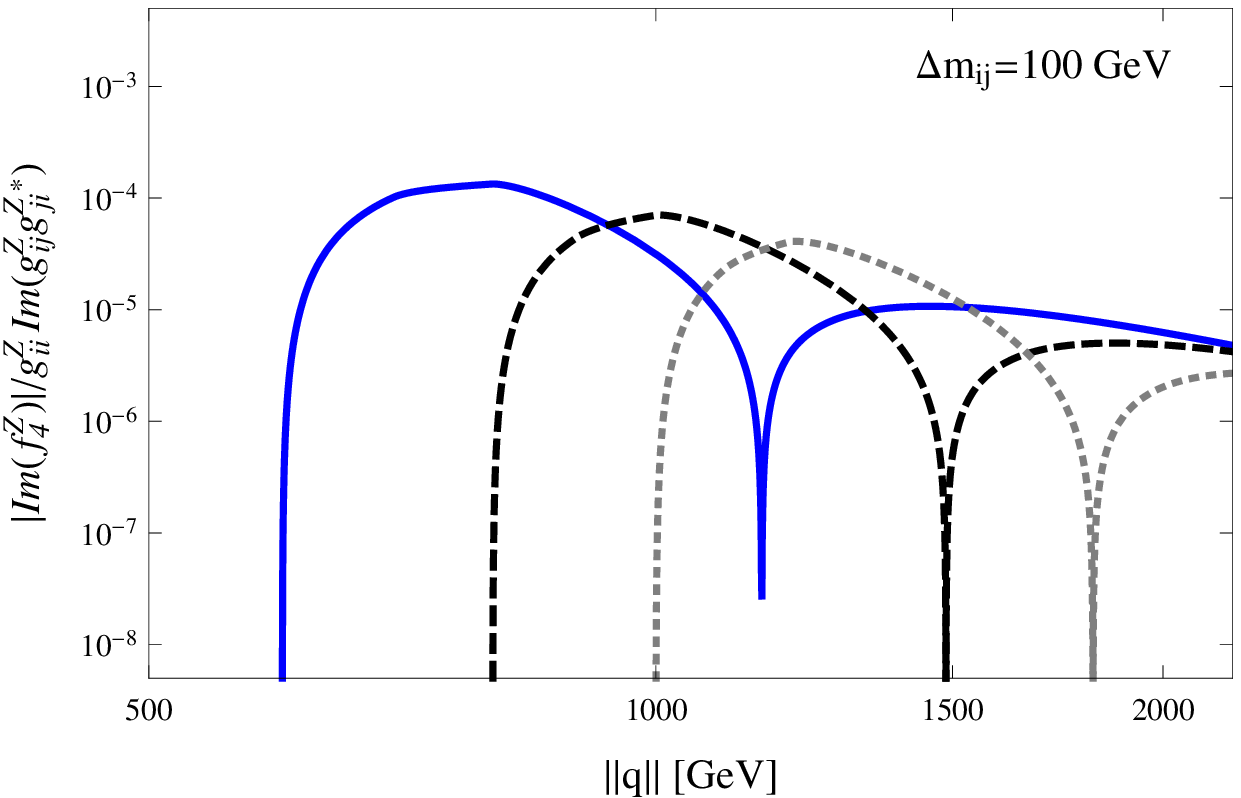}
\includegraphics[width=8cm]{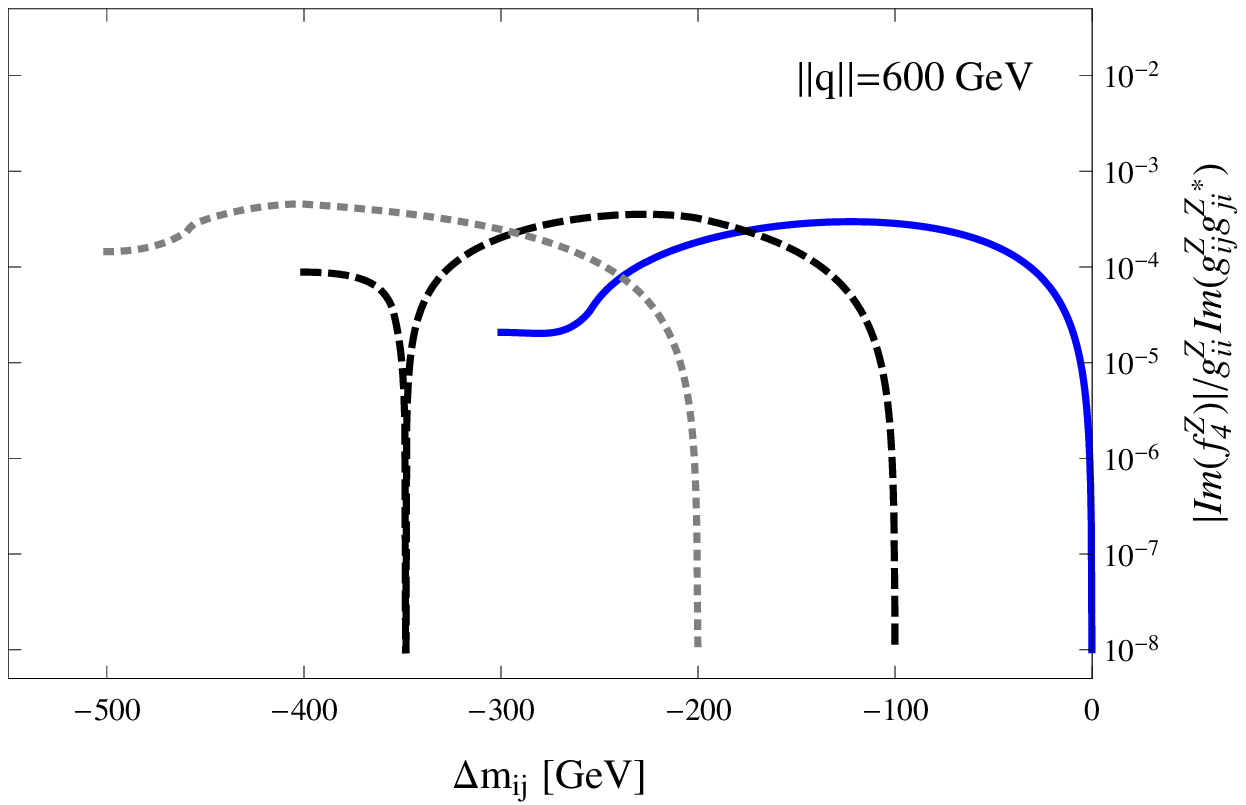}
\caption{The same as in Fig. \ref{f4g} but for the $f_4^Z$ form factor. Unlike the $f_4^\gamma$
form factor, which becomes undefined in $||q||=0$, the
$f_4^Z$ form factor gets undefined in $||q||=m_Z$.}
\label{f4Zfig}
\end{figure}

Finally we show in Figure \ref{f4Zfig}  the real (top plots)  and imaginary (bottom plots) parts
of the form factor $f_4^Z$ as a function of the four-momentum
magnitude $||q||$ (left plots) and the mass splitting $\Delta m_{ij}$ (right plots). We consider the same scenarios analyzed above.

We first analyze the behavior of the real part of  $f_4^Z$ (top plots) as a function of  $||q||$
and $\Delta m_{ij}$. This form factor has a sharp peak at $||q||=m_Z$ due to the factor $q^2-m_Z^2$ in the denominator. There are also two dips appearing in the curves of the top-left plot, which as in the previous cases
are due to a sign flip of this form factor. We observe that,
unlike  the $h_{1,2}^Z$ form factors,  $f_4^Z$ flips sign twice for $m_Z\le ||q||\le 2000$ GeV. One of such sign flips occurs at
$||q||\simeq 110$ GeV, regardless of the  $m_i$ value, and  the second flip occurs at
$||q||\simeq 800$, 1000, and 1250  GeV for $m_i=300$, 400, and 500 GeV, respectively. Inside the
region enclosed by the two dips the real part of
$f_4^Z$   is positive, and it is negative outside this region. Although the larger values of $f_1^Z$ are reached around the $Z$ resonance, there is an increase of the real part of $f_4^Z$ in the interval between $||q||=$140 GeV and  $||q||=$700 GeV, where  the real part of $f_4^Z$ goes from  $10^{-6}$ up to $10^{-5}$, in units of $g_{ii}^Z
\text{Im}(g_{ij}^{Z}{g_{ji}^Z}^*)$.

The $f_4^Z$ for factor  develops an imaginary part when $||q||\ge\min
\left(2m_i,m_i+m_j,2m_j\right)$ since in this  loop the virtual $Z$ boson can now couple to the same or distinct
charged scalar bosons. In the bottom-left plot we have $m_j>m_i$, thus the imaginary part of $f_4^Z$
arises when $||q||\ge 2 m_i$. On the other hand, for the chosen values of $m_i$, in the left plot
we have $m_i\ge \frac{||q||}{2}$, thus the interval where the imaginary part of  $f_4^Z$ is nonzero
is given by $||q||\ge 2m_j$ or in terms of the mass-splitting $-m_i\le \Delta  m_{ij}\le \frac{||q||}{2}-m_i$, as observed in the curves shown in the bottom-right plot. Thus $f_4^\gamma$ and $f_4^Z$ develop imaginary parts in the same interval.
Most part of this interval correspond to a very light charged scalar and as occurs with the other form factors,   the imaginary part of $f_4^Z$  has a size of the
same order of magnitude than its  real part, which however can reach much higher values outside
this region.

It is interesting to note that the $f_4^Z$
form factor was also studied in the framework of a THDM where the respective contribution is induced by three nondegenerate  neutral scalar
bosons
\cite{Chang:1994cs}. In such a model three different nondiagonal complex couplings ${\mathcal O}_{1i}$ arise in the neutral
scalar sector. It  was reported in
\cite{Chang:1994cs} that $f_4^Z$ can reach values from $10^{-6}\times {\mathcal
O}_{11}{\mathcal O}_{12}{\mathcal O}_{13}$ to $10^{-5} \times {\mathcal O}_{11}{\mathcal
O}_{12}{\mathcal O}_{13}$, where the following set of values for the free parameters was used: $\sqrt{s}=||q||=200$ GeV, $M_3=250$ GeV,
$M_2=150$ GeV, and  60 GeV $\le M_1\le $ 150 GeV, with $M_i$   the masses of the neutral scalar
bosons. In this case the ${\mathcal O}_{11}{\mathcal O}_{12}{\mathcal O}_{13}$ factor could suppress
considerably the THDM contribution to $f_4^Z$, just as happens with the contribution of our charged scalar bosons, which could be suppressed by the factor $g_{ii}^Z
\text{Im}(g_{ij}^{Z}{g_{ji}^Z}^*)$.

\section{Conclusions }
We have presented an analysis of the one-loop contributions to the CP-violating form factors associated with the $ZZ^*\gamma$,
$ZZ\gamma^*$ and $ZZZ^*$ couplings   in the  framework of an arbitrary effective model with
at least two nondegenerate charged scalar bosons that couple nondiagonally to the $Z$ gauge boson.
Such form factors are induced only when the nondiagonal  coupling constants  $g_{ij}^Z$ are complex and  have
an imaginary phase.  Our analysis is independent of
any specific value of the coupling constants, so our results are scaled by the coefficient   $g_{ii}^V
\text{Im}(g_{ij}^{Z}{g_{ji}^Z}^*)$. We considered a charged scalar boson with  mass above 300
GeV, which is consistent with experimental constraints, and analyze the behavior of the real and imaginary parts of the form factors as  functions of the four-momentum of the virtual gauge boson $||p||$ or $||q||$ and the splitting of the masses of the charged scalar bosons $\Delta m_{ij}=m_j-m_i$.
Although the region which is consistent with experimental data corresponds to $\Delta m_{ij}\ge 0$, for completeness we also include  the results for $-m_i\le \Delta m_{ij}\le 0$, which corresponds to  $m_j$ lighter than $m_i$ . As far as the orders of magnitude  of the form factors are
concerned,  they  are as follows in units of $g_{ii}^V
\text{Im}(g_{ij}^{Z}{g_{ji}^Z}^*)$: $|h_1^Z|\sim 10^{-5}-10^{-4}$,  $|h_2^Z|\sim
10^{-7}-10^{-6}$, $|f_4^\gamma|\sim 10^{-5}-10^{-3}$ and  $|f_4^Z|\sim 10^{-6}-10^{-5}$, for values of $||p||$ about a few hundreds of GeVs.  When the magnitude of $||p||$ increases, the real part of the form factors decreases smoothly  and   decouple  at high
energy. Furthermore,  it is not very sensitive to
large values of $\Delta m_{ij}$.  As for its  imaginary part, for an intermediate value of $||p||$ it can arise in a small region of values of $m_i$ and $m_j$, which corresponds mainly to the scenario with a relatively light charged scalar. Such scenario is not favored by current constraints on the mass of a charged scalar boson. On the other hand, for heavy $m_i$ and $m_j$, the imaginary part of the form factors can only arise for large $||p||$. Therefore a higher energy would be required in order that the form factors could develop an imaginary part, which can have a magnitude similar to the corresponding real part, though  the latter can reach larger values in the region where the former vanishes.
We also find that, except for a proportionality factor, our results for the $f_4^Z$ form factor are of the same order of
magnitude than the contributions arising in a THDM with  three nondegenerate neutral scalar  bosons
that couple nondiagonally to the $Z$ gauge boson. It is worth noticing that our results are valid
for the contribution of nondegenerate doubly charged scalar bosons, though in this case the form
factors associated with the $ZZ ^*\gamma$ and $ZZ\gamma^*$ couplings would get an additional factor of
two.

Although the LHC was down temporarily, it has resumed their operations with a higher center of mass energy and a higher integrated luminosity. A  significant improvement in the experimental limits of the TNGBCs is thus expected in the forthcoming years. Therefore it is necessary to examine any potential contribution to the respective form factors.

\acknowledgments{We acknowledge support from Conacyt and SNI (M\'exico). Partial support from VIEP-BUAP is also acknowledge. }

\bibliography{article}{}

\end{document}